\documentclass[10pt,aps,prd,nofootinbib,reprint,preprintnumbers]{revtex4-1}

\usepackage[utf8]{inputenc}
\usepackage{amsmath}
\usepackage{graphicx}
\usepackage{multirow}
\usepackage{pgfornament}
\usepackage{array}

\usepackage[colorlinks,pdfusetitle]{hyperref}
\hypersetup{colorlinks=true,allcolors=[rgb]{1,0.56,0}}

\newcommand{\orcid}[1]{\href{https://orcid.org/#1}{#1}}
\newcommand{\e}[1]{\times10^{#1}}

\begin{document}

\title{Neutrino Constraints and the ATOMKI X17 Anomaly}

\author{Peter B.~Denton}
\email{pdenton@bnl.gov}
\thanks{\orcid{0000-0002-5209-872X}}
\affiliation{High Energy Theory Group, Physics Department, Brookhaven National Laboratory, Upton, NY 11973, USA}

\author{Julia Gehrlein}
\email{julia.gehrlein@cern.ch}
\thanks{\orcid{0000-0002-1235-0505}}
\affiliation{Theoretical Physics Department, CERN, 1211 Geneva 23, Switzerland}

\preprint{CERN-TH-2023-053}

\begin{abstract}
Recent data from the ATOMKI group continues to confirm their claim of the existence of a new $\sim17$ MeV particle.
We review and numerically analyze the data and then put into context constraints from other experiments, notably neutrino scattering experiments such as the latest reactor anti-neutrino coherent elastic neutrino nucleus scattering data and unitarity constraints from solar neutrino observations.
We show that minimal scenarios are disfavored and discuss the model requirements to evade these constraints.
\end{abstract}

\date{April 20, 2023}
\date{\today}

\maketitle

\section{Introduction}
Understanding the validity and meaning of any particle physics anomaly requires a careful understanding of the data pointing towards the anomaly, studies of the new physics scenarios compatible with the anomaly, a confirmation that any physics scenario is consistent with all other data, and finally predictions for upcoming experiments all within a statistical framework.
In the following, we will focus on data from the ATOMKI collaboration which has reported evidence for an anomaly in a suite of measurements looking at the angular distributions of the decays of excited light nuclei to $e^+e^-$, each of which is individually preferred over the Standard Model (SM) at $>5\sigma$ \cite{Krasznahorkay:2015iga,Krasznahorkay:2019lyl,Krasznahorkay:2021joi,Krasznahorkay:2022pxs}; for a recent summary of the status see \cite{Alves:2023ree}.
In \cite{Zhang:2017zap,Koch:2020ouk} nuclear physics explanations of the anomaly have been put forward; however, an explanation due to unknown nuclear physics has been deemed to be unlikely strengthening the case for a particle physics explanation of the data. Similarly, explanations within the Standard Model based on the presence of new exotic QCD states  \cite{Chen:2020arr,Kubarovsky:2022zxm,Wong:2022kyg} or so far unaccounted for Standard Model effects \cite{Aleksejevs:2021zjw,Hayes:2021hin,Viviani:2021stx} have up to now also not led to a conclusive explanation of the ATOMKI data. Therefore we turn our focus to explanations beyond the Standard Model; indeed all the data seems to be pointing to a new state with a mass of about 17 MeV based on a straightforward examination of the kinematics of the data.

The validity of the anomaly and the nature of the state is not yet fully understood.
Nonetheless, some facts about it seem to be increasingly clear.
After careful analyses of a variety of scenarios, the data seems to prefer a vector mediator \cite{Feng:2016ysn,Feng:2016jff,Pulice:2019xel,Feng:2020mbt,Nomura:2020kcw}, although an axial-vector mediator may also be allowed, depending on the exact treatment of other data sets and our understanding of nuclear  physics \cite{Kozaczuk:2016nma,DelleRose:2018pgm,Feng:2020mbt,Seto:2020jal,Barducci:2022lqd}.
Some analyses found that the angular distribution did not exactly match the vector boson solution \cite{Zhang:2020ukq}, although with more data from ATOMKI the situation becomes more unclear \cite{Sas:2022pgm,Alves:2023ree}.
In addition, some early analyses found preference for protophobic structures \cite{Feng:2016ysn}, however this statement will be reexamined here.

Such an MeV scale boson can be probed in neutrino scattering experiments, notably via the coherent elastic neutrino nucleus scattering (CEvNS) process \cite{Freedman:1973yd}.
In fact, CEvNS experiments provide strong bounds on new light mediators which couple to neutrinos and neutrons \cite{Coloma:2017egw,Liao:2017uzy,Coloma:2017ncl,Denton:2018xmq,Denton:2020hop,AristizabalSierra:2022axl,Liao:2022hno,Denton:2022nol,AtzoriCorona:2022moj}.
Crucial constraints on a 17 MeV mediator will come from reactor CEvNS experiments, at which there has not yet been a definitive detection.
Nonetheless, several experiments have limits very close to the expected signal which are enough to constrain relevant parameter space.
Recently several reactor CEvNS experiments have reported constraints close enough to the SM prediction to derive key constraints on the coupling of light mediators to nucleons and neutrinos \cite{Colaresi:2021kus,Colaresi:2022obx,CONUS:2020skt,CONUS:2021dwh,conus_m7}.
In the following, we will use this new data to constrain explanations of ATOMKI which we will show provides important requirements on complete descriptions of the anomaly.

We perform a new statistical analysis of parameters preferred by the latest ATOMKI data in the context of the vector mediator solution in section \ref{sec:atomki}.
In particular, we examine the self-consistency of the data under different assumptions about the nuclear physics.
We then discuss the generic constraints on such a scenario including the latest neutrino data from reactor CEvNS experiments in section \ref{sec:con_dom}.
We then turn to model specifics with an aim of understanding the minimal particle content required to explain the ATOMKI data beyond a new spin-1 boson at $\sim17$ MeV in section \ref{sec:scenarios}.
We discuss future tests of the anomaly and conclude in section \ref{sec:conclusion}.

\section{ATOMKI Hints for New Physics}
\label{sec:atomki}
Over the last several years, the ATOMKI collaboration reported several statistically significant excesses in the opening angle distributions of $e^+e^-$ pairs produced in the decays of excited states of Be \cite{Krasznahorkay:2015iga}, He \cite{Krasznahorkay:2019lyl,Krasznahorkay:2021joi}, and C \cite{Krasznahorkay:2022pxs} with multiple individual significances of $>5\sigma$ each.
These results have been interpreted as a hint for a new boson coupling to nucleons and electrons with a mass of $m_X\approx 17$ MeV.
Previous studies of the anomaly in Be and He showed that it is difficult to simultaneously explain these results with a scalar or pseudoscalar boson \cite{Feng:2020mbt}.
An axial vector solution benefits from avoiding the strong constraint on its coupling to protons from $\pi^0$ decays, but struggles due to large theory uncertainties \cite{Feng:2020mbt}, although see also \cite{Barducci:2022lqd}.
In any case, we show that these  constraints, when considered numerically along with the ATOMKI data, are not as limiting as previously thought.
Therefore we will focus in the following on a vector boson solution.
We consider a model with several free parameters, some to be constrained by the details of the production of  $X$ and others from the necessary decay requirements.
Constraints and preferred values on these parameters from other experiments will be considered in the next section.

The Lagrangian of the model reads
\begin{equation}
\mathcal{L}\supset \text{i}X_\mu e\varepsilon_i \overline{f}_i\gamma^\mu f_i\,,
\end{equation}
where $X_\mu$ is a new vector field which couples with coupling strength $\varepsilon_i$ to fermions $i$, $i=n,~p,~e,~\nu_e$ as minimally required by the ATOMKI data and $e$ is the elementary charge.

\begin{figure}
\centering
\includegraphics[width=\columnwidth]{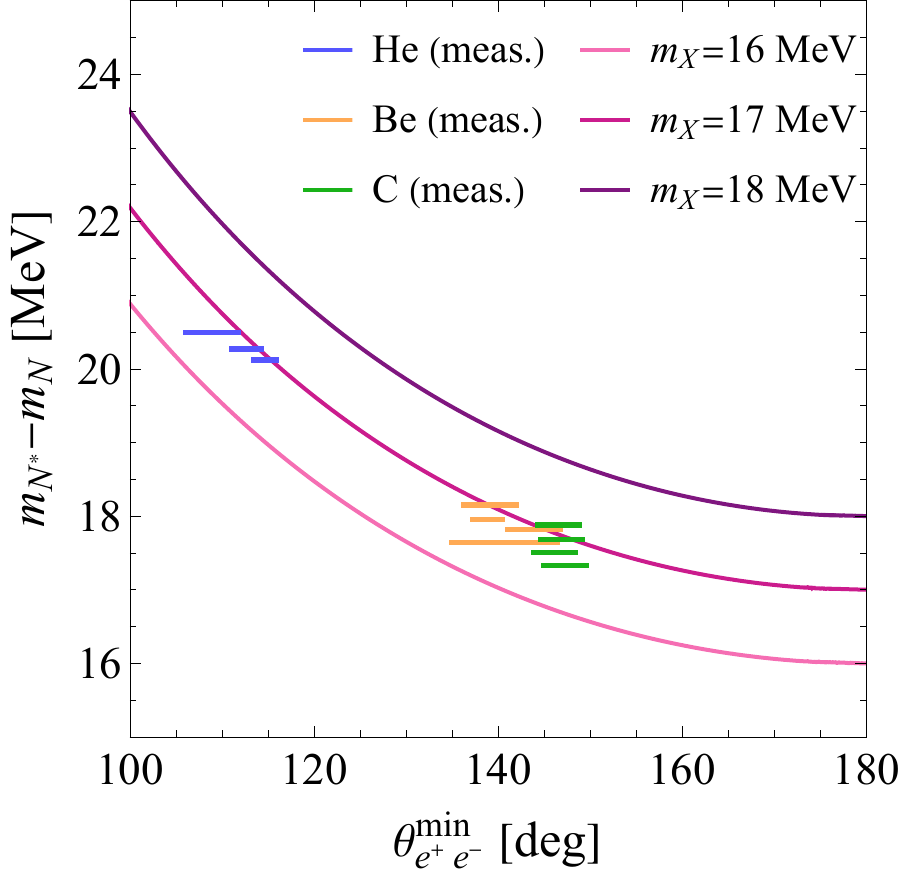}
\caption{Measured opening angles of the $e^+e^-$ pairs using the mass differences between different excited states and the ground state of  He (blue), Be (orange), C (green).
We show contours of different $m_X$ using the relation $\theta_{e^+e^-}^{\text{min}}\approx 2\arcsin(m_x/(m_{N^*}-m_N))$ \cite{Feng:2020mbt}.}
\label{fig:resultangle}
\end{figure}

\begin{figure*}
\centering
\includegraphics[width=0.49\textwidth]{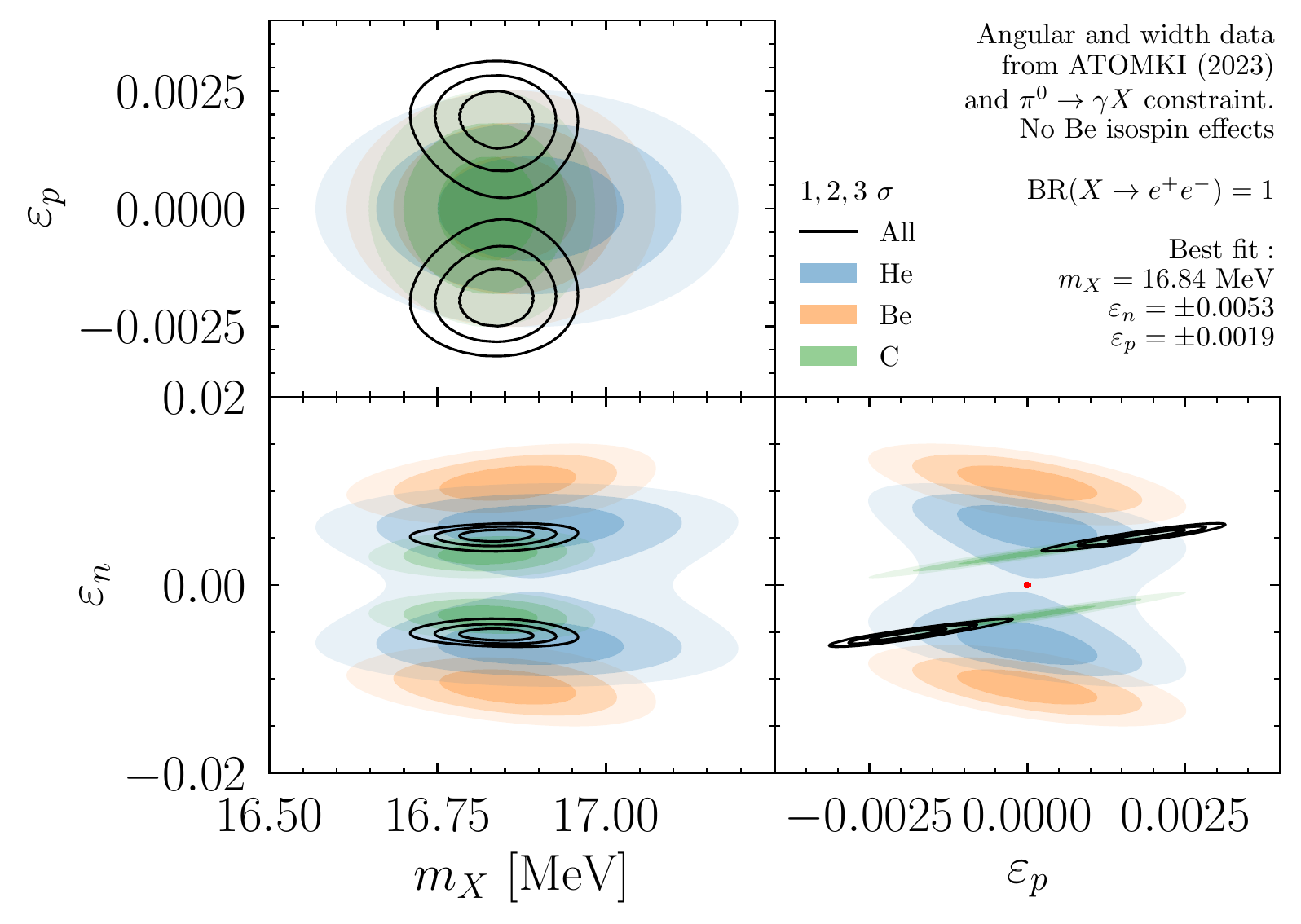}
\includegraphics[width=0.49\textwidth]{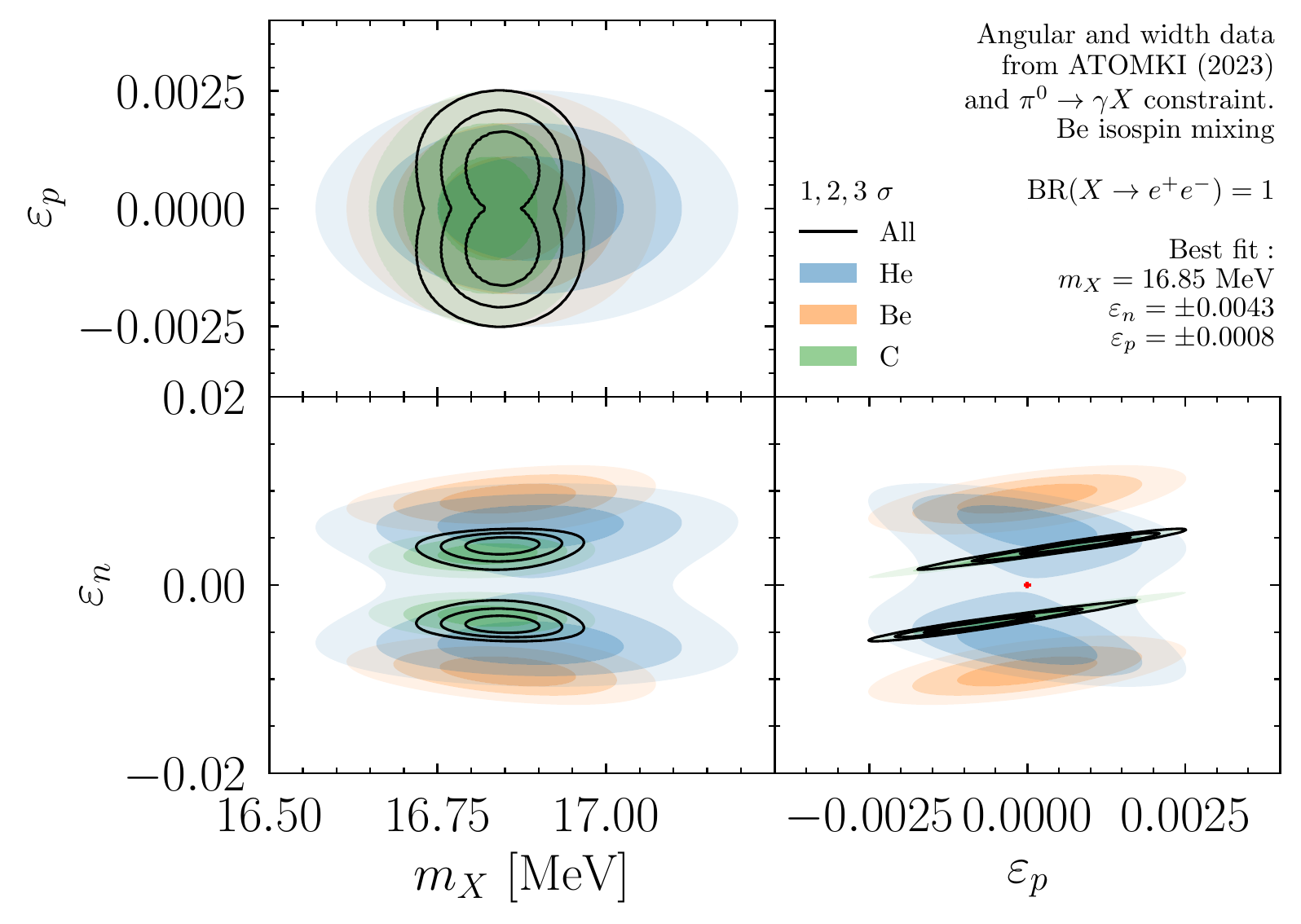}
\includegraphics[width=0.49\textwidth]{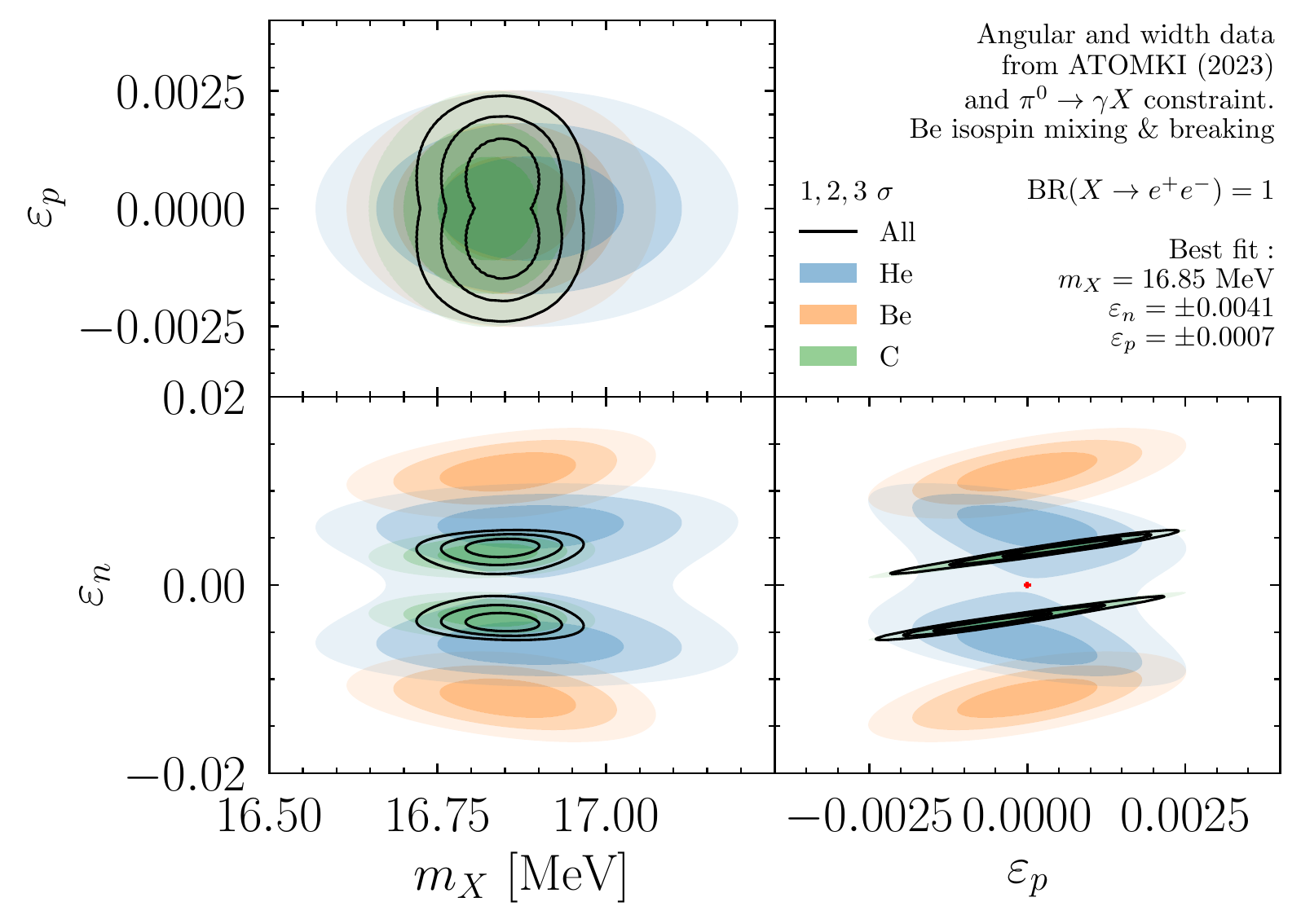}
\caption{The parameter estimation at 1, 2, 3$\sigma$ of $m_X$, $\varepsilon_n$, and $\varepsilon_p$ using 11 separate angular measurements and the three latest width measurements from ATOMKI in addition to a prior on $\varepsilon_p$ from $\pi^0\to\gamma X$ constraints for three separate treatments of the Be nuclear physics: no isospin effects (top left), isospin mixing (top right), and both isosopin mixing and breaking (bottom).
The not shown parameter is minimized over in each panel.
The colors correspond to the preferred parameters from individual ATOMKI measurements and the black curves are the result of the combined fit.
We assume that BR($X\to e^+e^-$)${}=1$.
}
\label{fig:angular width}
\end{figure*}

We now turning to our numerical analysis.
The ATOMKI data is compelling because there is a fairly self-consistent picture of new physics at $\sim17$ MeV coupling to protons and/or neutrons and electrons from data from different angular distributions, widths, and elements. The ATOMKI data comes in two dimensions: the angle at which the $e^+e^-$ excess over the background begins, and the rate leading to the excess.
These can be parameterized in the quantities $\theta_{e^+e^-}^{\min}$ and $\Gamma_X/\Gamma_\gamma$ where the second parameter is the ratio of partial widths to the new $X$ boson and to a photon which is both experimentally and theoretically convenient ratio to take\footnote{Note that this ratio is often confusingly referred to as a ``branching ratio.'' Since $\Gamma_\gamma\ll1$ for all three elements, this ratio is quite different from the branching ratio to $X$.}.
We use the calculations of the kinematics for the angle and the widths in the vector case from \cite{Feng:2020mbt} to compare with the data.

For the angular data we use the data from \cite{Sas:2022pgm,Krasznahorkay:2021joi,Krasznahorkay:2022pxs} as extracted in \cite{VerhaarenSlides} including three measurements with He, four measurements with Be, and four measurements with C, see fig.~\ref{fig:resultangle}.
For the width data we use \cite{Firak:2020eil} for the Be data, \cite{Krasznahorkay:2022pxs} for the C data, and \cite{Krasznahorkay:2021joi} for the He data.
For the He data we also include the theory uncertainty on $\Gamma_{E0}$ \cite{Walcher:1970vkv}, the width normalization used for He coming from the $0^+\to0^+$ transition.

We perform a simple statistical $\chi^2$ test of all the data from multiple experimental runs of each of the three elements including width and angular information to compute the preferred parameters and the internal goodness-of-fit of the model using the procedure outlined in \cite{Feng:2020mbt}.
We do not perform a model comparison test between new physics and the Standard Model as this requires more intimate knowledge of the experimental details and since new physics is preferred over no new physics at very high significance $\gg5\sigma$.

An analysis with the angular data alone of 11 different measurements finds that the data is well described by a new particle of mass $m_X=16.85\pm0.04$ MeV with an internal goodness-of-fit of $1.8\sigma$ calculated from Wilks' theorem at $\chi^2/dof=17.3/10$.
We use only the best fit and uncertainty of the maximum of the angular distribution; a more complete angular distribution might slightly modify the results due to fluctuations in the data.
The data is compatible with the expected signature from a $\sim17$ MeV mediator, so we find it unlikely that this will significantly shift the results.
The angular distributions are only sensitive to the mass of the particle which makes it a useful starting point in analyzing the ATOMKI measurements.

Next, we add in to the analysis the latest width information from each element and include a prior on $\varepsilon_p$ since $X$ needs to couple to protons and/or neutrons on the production size.
There is a stronger constraint on the coupling of $X$ to protons from measurements of $\pi^0$ decays than the constraint on the coupling to neutrons.
We will include a prior on the coupling to protons $|\varepsilon_p|\lesssim1.2\e{-3}/\sqrt{\text{Br}(X\to e^+e^-)}$ at 90\% \cite{Raggi:2015noa,Feng:2016ysn}; see the next section for more information.

 For the Be scattering, it is possible that one of the states has isospin breaking effects which complicates the picture \cite{Feng:2016ysn,Feng:2020mbt}.
 As such, we present our results for three different possible interpretations of the nuclear physics: that without any modifications due to isosopin (top left of fig.~\ref{fig:angular width}), that with isospin mixing (top right), and that with both isospin mixing and breaking (bottom).

\begin{table}
\centering
\caption{The preferred values for the three different treatments of isospin in the Be scattering as well as the internal goodness-of-fits using all ATOMKI data and the $\pi^0$ constraint.
The signs of the couplings are correlated.}
\begin{tabular}{l||c|c|c|c}
 & $m_X$ (MeV) & $\varepsilon_n$ & $\varepsilon_p$ & $N\sigma$\\\hline\hline
No isospin effects & 16.84 & $\pm0.0053$ & $\pm0.0019$ & 3.7\\\hline
Isospin mixing & 16.85 & $\pm0.0043$ & $\pm0.0008$ & 4.6\\\hline
Isospin mixing & \multirow2*{16.85} & \multirow2*{$\pm0.0041$} & \multirow2*{$\pm0.0007$} & \multirow2*{5.0}\\[-0.1ex]
and breaking &&&
\end{tabular}
\label{tab:results}
\end{table}

For the scenario with no isospin effects in the Be system, we find that there is modest internal tension in the data and if the isospin effects are important than the internal goodness-of-fit of the scenario is poor, as shown in fig.~\ref{fig:angular width} and table \ref{tab:results}.
We note that the signs of $\varepsilon_n$ and $\varepsilon_p$ must be the same due to the non-trivial degeneracy structure shown clearly in the $\varepsilon_n$ -- $\varepsilon_p$ plots in all three panels of of fig.~\ref{fig:angular width}.
We have confirmed that the mass constraint is dominated by the angular data and is only weakly affected by the width data.
The internal goodness-of-fits range from $3.7\sigma$ (no isospin effects) to $5.0\sigma$ (both isospin mixing and breaking) indicating somewhere between modest and significant tension in the data within the explanation using a vector boson.
Thus if this new physics exists, it might suggest that in reality there are none or fewer isospin effects in the Be system, although such a conclusion would require confirming the new physics scenario in other environments as well as independent tests of the nuclear physics.
Alternatively, it could be that there is an issue with either the Be or C data sets, as these are the two data sets driving the internal tension.

We see that the preferred value of $|\varepsilon_p|\in\{0.7,1.9\}\e{-3}$ may be pulled somewhat above  the existing $90\%$ limit of $1.2\e{-3}$ in some cases.
Nevertheless, this contributes no more than 7 units of $\chi^2$ to the goodness-of-fit while the biggest part comes from the disagreement between the preferred values from the Be and C data.
The data prefers this somewhat larger value of $|\varepsilon_p|$ in the scenario without isospin effects because the rate measured by ATOMKI with carbon is lower than would be expected from the Be and He measurements if $\varepsilon_p=0$.
This difference can be partially accommodated because the widths for Be and He are proportional to $(\varepsilon_n+\varepsilon_p)^2$ while the width for C is proportional to $(\varepsilon_n-\varepsilon_p)^2$ and thus the inclusion of non-zero $\varepsilon_p$ leads to a partial cancellation reducing the C rate, while it enhances the rates for Be and He.
When including isospin effects, however, the width of Be is dominantly proportional to $(\varepsilon_p-\varepsilon_n)^2$ and therefore the Be and C preferred regions are relatively parallel to each other and do not significantly overlap.

To summarize our analysis of the ATOMKI data, we find that the data is in excellent agreement on the mass of the mediator which is dominated by the angular data.
The rate measurements which provide the information about the couplings $\varepsilon_p$ and $\varepsilon_n$ are not in agreement, although the size of the tension depends on uncertain isospin effects and is smaller than the overall evidence for new physics.

\section{Constraints}
\label{sec:con_dom}
The interactions of a new mediator with $\mathcal{O}$(MeV) mass scale can be probed with low-energy  experiments. Below we summarize the dominant constraints on the couplings of a vector boson $X$; appendix \ref{sec:constraints} contains further sub-dominant constraints coming from electron-neutrino scattering, invisible decays of $X$, and the lifetime of $X$.

As briefly mentioned in the previous section, constraints on the couplings of $X$ to quarks come from the search for rare pion decays $\pi^0\to \gamma X,~X\to e^+e^-$ where NA48/2 provides currently the strongest bound \cite{Raggi:2015noa}.
We follow \cite{Feng:2016ysn} to translate the bound to obtain the bound on the coupling to protons $|2\varepsilon_u+\varepsilon_d|=|\varepsilon_p|<((0.8-1.2)\times 10^{-3})/\sqrt{\text{BR}(X\to e^+e^-)} $ at 90\% C.L.~where the range in the constraint comes from the fast oscillating nature of the bounds around $m_X=17$ MeV.
We somewhat optimistically  take the $1.2\e{-3}$ number as the ATOMKI data may prefer $|\varepsilon_p|$ on the larger side.
We note that the coupling to protons is really a combination of the couplings to up and down quarks which will be discussed in further detail below.

Since $X$ must decay to $e^+e^-$, it must also couple to electrons.
A new light mediator coupling to electrons leads to a contribution to the electron $g_e-2$ \cite{Leveille:1977rc}. Recently a new measurement of this quantity has been reported \cite{Fan:2022eto} which deviates from the SM expectation using the measured value of the electromagnetic fine structure constant by $\sim 3\sigma$
\cite{Morel:2020dww,Parker:2018vye}\footnote{Note that there are two independent measurements of the fine structure constant  which disagree at $5.4\sigma$ \cite{Morel:2020dww,Parker:2018vye}.}. Using the SM prediction from \cite{Morel:2020dww}, this discrepancy leads to a mild preference for a new mediator with $\varepsilon_e=(7.0\pm1.5)\times 10^{-4}$ at $m_X=17$ MeV  but also disfavors $\varepsilon_e>1.2\times 10^{-3}$ at 90\% C.L.

A lower limit on the coupling of $X$ to electrons comes from searches using the bremsstrahlung reaction $e^-Z\to e^-Z X$ and the subsequent decay of $X$  into an electron-positron pair.
From the null results of this search at the NA64 experiment \cite{NA64:2019auh} we get $|\varepsilon_e|> (6.3\times10^{-4})/\sqrt{{\text{BR}}(X\to e^+e^-)}$ for $m_X=17$ MeV at 90\% C.L.
Combined with the $g_e-2$ constraint this leads to an allowed range of $\varepsilon_e \in [0.63,1.2]\times 10^{-3}$ for ${\text{BR}}(X\to e^+e^-)=1$. For smaller electron couplings  $X$ escapes the detector and no bounds can be derived at terrestrial experiments  ($|\varepsilon_e|<10^{-7}/\sqrt{{\text{BR}}(X\to e^+e^-)}$ \cite{Bjorken:1988as}).
However from the absence of electromagnetic signals from the decay of a dark photon near the surface of a supernova progenitor star\footnote{Other analyses \cite{Bjorken:2009mm,Dent:2012mx,Dreiner:2013mua,Chang:2018rso} do not include this effect and find constraints weaker by two orders of magnitude at 17 MeV. Even if this effect is not included, neither our discussion here nor the unitarity issue presented later change.} we get the very strong constraint $|\varepsilon_e|<10^{-12}/\sqrt{{\text{BR}}(X\to e^+e^-)}$ \cite{Kazanas:2014mca} leading to second allowed region for small electron couplings of $X$.

Depending on the details of the model, the mediator may well couple to neutrinos in addition to electrons. Note that a vector mediator  which couples to charged leptons automatically also couples to neutrinos.
A new light mediator which couples to neutrinos and neutrons is constrained by CEvNS with reactor neutrinos \cite{Colaresi:2022obx,AristizabalSierra:2022axl,Denton:2022nol,Coloma:2022avw}\footnote{Note that the constraints from reactor experiments only apply to the coupling with electron neutrinos. Nevertheless, the constraints on coupling to muon neutrinos are only slightly less stringent  while the coupling to tau neutrinos is about an order of magnitude less constrained \cite{Kling:2020iar}.}.
From Dresden-II data, we obtain as constraints
at $m_X=17$ MeV, $\sqrt{|\varepsilon_n\varepsilon_{\nu_e}|}<13.6~ \times 10^{-5}$ for $\varepsilon_n\varepsilon_{\nu_e}>0$, $\sqrt{|\varepsilon_n\varepsilon_{\nu_e}|}<8.0~ \times 10^{-5}$ for $\varepsilon_n\varepsilon_{\nu_e}<0 $ at 90\% C.L. 
Similar constraints exist from COHERENT and CONUS \cite{AristizabalSierra:2022axl,Coloma:2022avw,AtzoriCorona:2022moj,CONUS:2021dwh,CONNIE:2019xid}.
Since the ATOMKI data is equally explained for nucleon couplings of either sign, we take the more conservative of the two constraints: the positive product constraint.
This then sets the constraint on the neutrino coupling that must be avoided.

To summarize this section, a model with a vector mediator explaining the ATOMKI anomaly at a minimum needs to fulfill the following requirements:
\begin{itemize}
\item feature a vector mediator with mass $m_X\approx 17$ MeV,
\item $X$ needs to couple to neutrons with strength \\$|\varepsilon_n|\sim(4.1-5.3)\e{-3}$,
\item $X$ needs to couple to protons with strength \\$|\varepsilon_p|\sim(0.7-1.9)\e{-3}$,
\item the product of neutron and proton couplings of $X$ need to fulfill $\varepsilon_n\varepsilon_p>0$,
\item the coupling of $X$ to electrons needs to be either $|\varepsilon_e|\in [0.63,1.2]\times 10^{-3}$  or $|\varepsilon_e|<10^{-12}$ \\for ${\text{BR}}(X\to e^+e^-)=1$, and
\item the coupling of $X$ to electron neutrinos needs to be smaller than $|\varepsilon_{\nu_e}|<(3.5-4.5)\e{-6}$.
\end{itemize}
The ranges in several of the bullet points covers the spread in the preferred values of the data depending on the proper treatment of the isospin effects in Be.

\begin{table}
\centering
\caption{The same as table \ref{tab:results} in terms of couplings to quarks.
The signs of the couplings are correlated.}
\begin{tabular}{l||c|c}
 & $\varepsilon_d$ & $\varepsilon_u$\\\hline\hline
No isospin effects & $\pm0.0029$ & $\mp0.0005$\\\hline
Isospin mixing & $\pm0.0026$ & $\mp0.0009$\\\hline
Isospin mixing & \multirow2*{$\pm0.0025$} & \multirow2*{$\mp0.0009$}\\[-0.1ex]
and breaking &&
\end{tabular}
\label{tab:results quarks}
\end{table}

While we have considered the constraints in terms of $\varepsilon_n$ and $\varepsilon_p$, we can recast the constraints in terms of up and down quarks as shown in table \ref{tab:results quarks}.
From these constraints we see that any model to explain the anomaly needs to violate $SU(2)_L$ invariance as the required couplings to electrons and neutrinos  do not follow the expectation $2\varepsilon_{\nu_e}=\varepsilon_e$.
Similarly the couplings to up and down quarks need to be unequal.
In fact, an ``upphobic'' ($\varepsilon_u=0$) scenario, where the coupling of $X$ to up quarks is suppressed, fits the data about as well as the general scenario.

Finally, a new mediator that explains the ATOMKI anomaly is only required to couple to first generation fermions; if it also couples to the other generation potentially more constraints need to be taken into account, see e.g.~\cite{Castro:2021gdf}.

The scenario with $\varepsilon_e=10^{-3}$ and $\varepsilon_{\nu_e}=4\e{-6}$  also leads to non-standard neutrino interactions (NSI) that affect neutrino oscillation experiments \cite{Wolfenstein:1977ue,Proceedings:2019qno}.
Given $\varepsilon_{\nu_e}\approx 4\e{-6}$ at the limit from CEvNS, we find that at $m_X=16.8$ MeV, the relevant NSI parameter is $\varepsilon_{ee}^d=\pm0.1$ which is currently allowed by fits to oscillation data \cite{Esteban:2018ppq}.
As the least constrained NSI parameter, improving constraints on $\varepsilon_{ee}$ is a top priority for oscillation experiments.
Future probes of NSIs by comparing \cite{Denton:2022nol} measurements of $\Delta m^2_{21}$ from JUNO \cite{JUNO:2022mxj} and with solar neutrinos at DUNE \cite{Capozzi:2018dat,DUNE:2020ypp} will be sensitive to $\varepsilon_{ee}^d=0.019$ at $1\sigma$ and thus provides a $\gtrsim5\sigma$ means of probing this scenario.
While COHERENT and other $\pi$-DAR CEvNS experiments lose sensitivity in this mediator mass range \cite{Abdullah:2022zue}, improved measurements of CEvNS with reactor neutrinos will improve upon these constraints as well \cite{Fernandez-Moroni:2021nap}.

\begin{figure}
\centering
\includegraphics[width=\columnwidth]{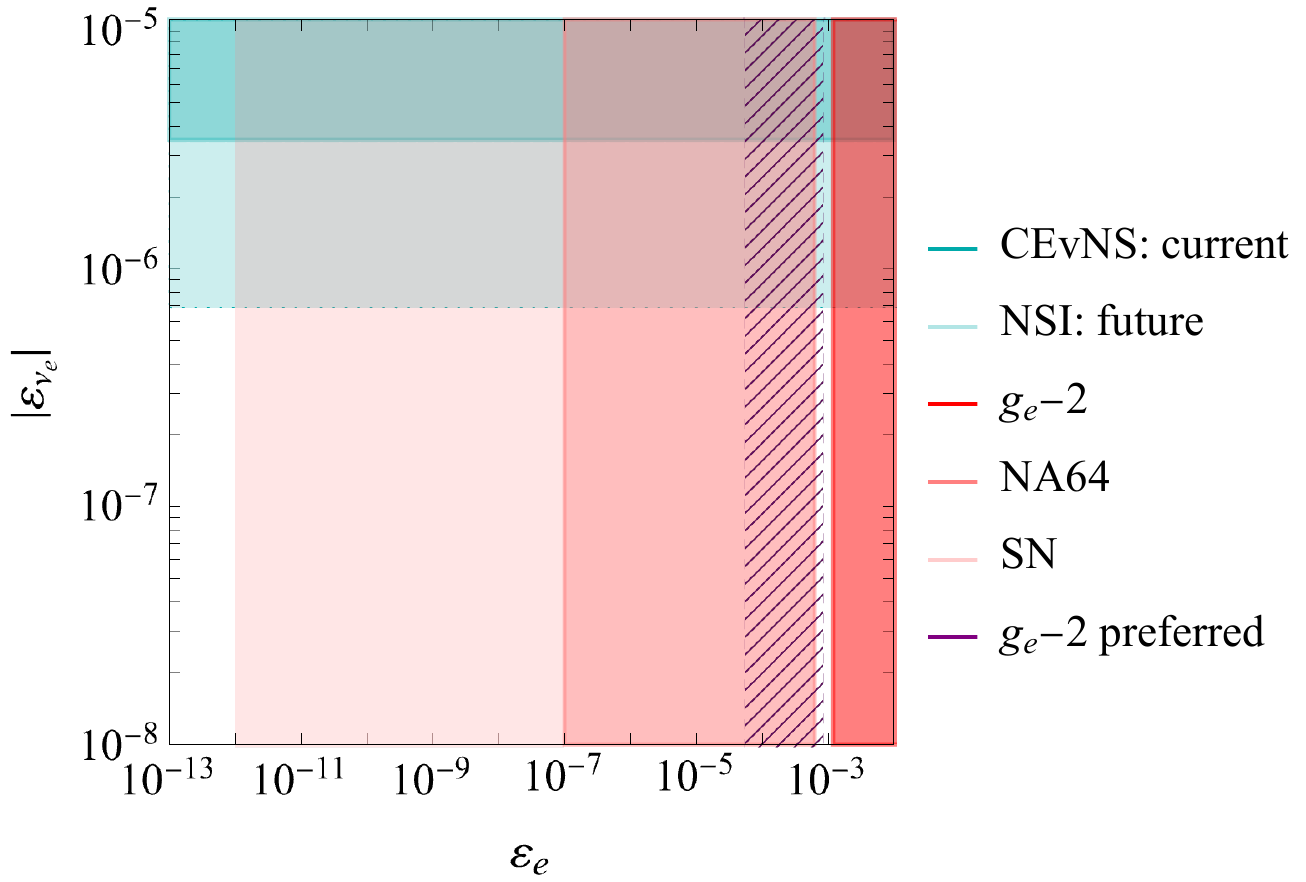}
\caption{Constraints on $\varepsilon_{\nu_e}$ and $|\varepsilon_e|$ for $m_X=17$ MeV. The dark cyan regions shows the current constraint from CEvNS setting $\varepsilon_n=0.0053$, the lighter cyan regions shows the future constraints from NSI at upcoming oscillation experiments. The red and pink regions show the excluded regions from NA64 and $g_e-2$, the lighter pink region to the left shows the constraint from SN. The purple hatched region shows the preferred region for $\varepsilon_e$ from $g_e-2$. The currently allowed region of parameter space is shown in white. The allowed region for $\varepsilon_e$ can be probed with future collider and beam dump experiments \cite{Ilten:2015hya,Apyan:2022tsd,FASER:2018eoc,Echenard:2014lma}.}
\label{fig:epsnuepse}
\end{figure}

Future probes of the parameter space will narrow it substantially down as shown in fig.~\ref{fig:epsnuepse} increasing the challenges of building a viable model. 
Therefore the experimental progress should also be accompanied by model building advances to find a viable model to explain the anomaly.

\section{Scenarios}
\label{sec:scenarios}
Following the general requirements on models to explain ATOMKI, we face potential model building challenges to realize small neutrino and up quark couplings while allowing for sizable electron and down quark couplings. Additional  model building complications could appear in the scenario without isospin effects of Be where the best fit value of the  proton coupling slightly exceeds the bounds from pion decays. Bringing the proton coupling in agreement with the constraint at 90\% leads to a mild decrease in the goodness-of-fit from $3.7\sigma$ to $4.0\sigma$.
While the constrain on the proton coupling can be evaded (and when including isospin effects for Be the best fit values of the proton couplings satisfy the bound) the requirement of a small neutrino coupling  is  independent\footnote{The upper bound on the neutrino coupling is affected by the isospin treatment but nevertheless, the neutrino coupling always needs to be much smaller than the neutron or electron coupling. } of the treatment of isospin effects of Be but merely  comes from the strong bounds from neutrino data and therefore it needs to be satisfied in all models which aim to explain the ATOMKI data. 
Thus, in the following we focus on model scenarios which achieve this feat. 
We split the scenarios into two main categories: those with large $\varepsilon_e\sim10^{-3}$ and those with small $\varepsilon_e\lesssim10^{-7}$.

\subsection{Large \texorpdfstring{$\varepsilon_e$}{varepsilon e} Scenarios}
We set $\varepsilon_e=10^{-3}$ in between the $g_e-2$ and NA64 constraints.
In fact, this region may be slightly preferred by the $g_e-2$ measurements and thus a discovery could be imminent by both $g_e-2$ measurements and NA64-like experiments.
This leads to BR($X\to e^+e^-$)${}=1$.
The BR to $e^+e^-$ is independent of the coupling to neutrons since we have $|\varepsilon_n|\sim(4.1-5.3)\e{-3}$ and thus the upper limit on $|\varepsilon_{\nu_e}|$ is $(3.5-4.5)\times10^{-6}$ from CEvNS at 90\% C.L.

\subsubsection{Flavor non-universal \texorpdfstring{$U(1)_X$}{U(1)X}  or anomalous \texorpdfstring{$U(1)_B$}{U(1)B}}
We are aware of two possible ways to proceed.
The first is to set $\varepsilon_\nu=0$ for example via
a flavor non-universal $U(1)_X$ model \cite{Pulice:2019xel}, where the charge of the first and second quark generations are identical and different from the ones of the third generation quarks while the lepton charges are universal. A charge assignment can be found which allows one to cancel all anomalies within the SM particle content and no new fermions need to be introduced.
In this model the new gauge boson mixes with the hypercharge gauge boson which  leads to $\varepsilon_{\nu_e}=0$. 

Alternatively, 
a $U(1)_B$ model could be introduced which is  however anomalous \cite{Feng:2016ysn}.
In this scenario the new boson (which will become the 17 MeV $X$ state) mixes with the photon allowing for different $\varepsilon_p$ and $\varepsilon_n$.
There is a body of literature on the additional particle content required to cancel anomalies; any such method can be applied \cite{FileviezPerez:2010gw,Dulaney:2010dj,FileviezPerez:2011dg,FileviezPerez:2011pt,Duerr:2013dza,Arnold:2013qja,FileviezPerez:2014lnj,Duerr:2014wra,Farzan:2016wym}.

\subsubsection{Anomaly Free \texorpdfstring{$U(1)_{B-L}$}{U(1)B-L}}
If one chooses to avoid an anomalous model but wants to make use of an accidental global symmetry of the SM like baryon number or $B-L$, we quantitatively present here an anomaly free model that allows for different $\varepsilon_p$ and $\varepsilon_n$ in the same way as in the $U(1)_B$ model.
To be concrete, we focus on a broken $U(1)_{B-L}$ model as described in \cite{Feng:2016ysn} which immediately leads to $\varepsilon_p\neq\varepsilon_n=-\varepsilon_{\nu_e}$ via a kinetic mixing between the new boson (which will become the 17 MeV $X$ state) and the photon as in the $U(1)_B$ model above.
We update this model to comply with additional neutrino constraints which leads to two possible ways of proceeding.
Mass in the dark sector is generated via a new $B-L$ Higgs boson with a vev of $3.4$ GeV that gives mass to $X$.

Since we need $|\varepsilon_\nu|$ much smaller than $\varepsilon_n$, we again follow \cite{Feng:2016ysn} with the suggested extension of including an additional vectorlike leptonic $SU(2)_L$ doublet.
After diagonalizing the mass matrix of the various neutrinos, we find that the remaining contribution to the neutrino coupling to $X$ is
\begin{equation}
\varepsilon_\nu=-\varepsilon_n\cos2\theta\,,
\label{eq:neutrino neutralization}
\end{equation}
where $\theta$ is the mixing between the active neutrino and the new vectorlike neutrino $\nu_4$.
Thus we require $|1-\tan\theta|<(7,11)\e{-4}$ to be consistent with neutrino scattering data which implies a fairly specific relation between seemingly unrelated parameters in the model.
The mixing angle depends on the number of new fermions and their masses.
For $N$ new neutrinos their masses must be given by this expression:
\begin{equation}
\sqrt{\tan\theta}=\left(\frac{70{\rm\ GeV}}{m_{\nu_4}}\right)\left(\frac{0.005}{|\varepsilon_n|}\right)\left(\frac{\sqrt N\lambda}{4\pi}\right)\simeq1\,,
\label{eq:tanphi}
\end{equation}
where the coupling $\lambda$ between the active neutrino and the $\nu_4$ state mediated by the new Higgs boson can be as large as $4\pi$.
Smaller values of $\lambda$ lead to smaller physical masses $m_{\nu_4}$.
This implies that we must have a new neutrino with a mass $\lesssim(70-90)$ GeV.

This new state cannot be lighter than $m_Z/2$ \cite{ParticleDataGroup:2022pth}, it must be heavier than $\sim50$ GeV, otherwise it would contribute to the well measured $Z$ width.
In addition, since the mixing angle with the light neutrino needs to be very close to $45^\circ$, this predicts very large unitary violation of the $\nu_e$ row of the measurable $3\times3$ PMNS matrix.
This can be constrained by comparing theoretical predictions for the reactor, solar, or radioactive source neutrino fluxes.
The measurement of $^7$Be neutrinos is in good agreement on the flux which, combined with shape information from KamLAND \cite{Bergstrom:2016cbh} provides a fairly direct constraint on the unitarity of the $\nu_e$ row at the few \% level.
Reactor neutrinos had a hint of a $\sim10\%$ deviation between the theoretical prediction and the measurement \cite{Mention:2011rk}, although careful measurements of the relative fluxes from different isotopes indicate that a nuclear physics issue may explain this tension \cite{DayaBay:2017jkb,RENO:2018pwo}.
Finally, there exists an unresolved tension in the comparison of the expected rate of neutrinos from $^{37}$Ar and $^{51}$Cr decays and the measurements \cite{GALLEX:1994rym,GALLEX:1997lja,Kaether:2010ag,Abdurashitov:1996dp,Giunti:2010zu,Kostensalo:2019vmv,Barinov:2021asz} which seems to predict quite large mixing at the $\sim40\%$ level, although in tension with solar results.

The strongest solar neutrino bound that directly contains the electron neutrino row normalization is the $^7$Be measurement which is in the low energy vacuum regime.
It is measured at the 8\% level at 90\% C.L.~and is consistent with the expectation at $<1\sigma$ \cite{Bergstrom:2016cbh}.
Since the $^7$Be is mostly the vacuum dominated regime, the probability, without assuming unitarity, is
\begin{equation}
P_{ee,D,vac}=(|U_{e1}|^2+|U_{e2}|^2+|U_{e3}|^2)^2-2|U_{e1}|^2|U_{e2}|^2\,,
\end{equation}
up to small $|U_{e3}|^2$ corrections.
Using the measurement from KamLAND and the theory prediction of the flux, this implies an uncertainty on the electron row normalization of 4\% at 90\% C.L., strongly disfavoring a maximal active-sterile mixing angle.
That is, at 90\% C.L.~the deviation is constrained to be $\delta_e\equiv1-(|U_{e1}|^2+|U_{e2}|^2+|U_{e3}|^2)<0.04$ from solar neutrino measurements\footnote{Constraints from fits to a large number of oscillation observables also exist \cite{Parke:2015goa,Ellis:2020hus,Hu:2020oba}, however these analyses are not truly global as they do not include all available data neither anomalous results.
In addition, some of the analyses assume that other experiments measured exactly the standard prediction, even when they did not.
Nonetheless, a more comprehensive analysis may well lead to stronger constraints on $\delta_e$ than that quoted here.}.

\textbf{Following the gallium anomaly}:
If we take the gallium anomaly's $>5\sigma$ hint of large unitarity violation in the $\nu_e$ row seriously, then the above model is valid and we are already seeing the large predicted unitarity violation.
The model also predicts a Majorana mass term for the SM singlet, right handed neutrinos which explains the mass of the active neutrinos through a seesaw via the same $B-L$ Higgs boson.
This Majorana mass needs to be $m_M\lesssim10$ GeV at the largest allowed coupling from unitarity to get the known active neutrino masses.
Thus we need a Dirac mass contribution of
\begin{equation}
m_D=14{\rm\ keV}\sqrt{\frac{m_M}{10{\rm\ GeV}}}\,,
\end{equation}
which gives a mass for the active neutrino of $m_{\nu_L}=0.01$ eV, and an additional neutrino exists at $m_{\nu_R}=m_M\lesssim10$ GeV.
Thus the active neutrino mixes with the right handed sterile neutrino at the level
\begin{equation}
\psi^2=1.4\e{-6}\sqrt{\frac{10{\rm\ GeV}}{m_M}}\,.
\end{equation}
At the maximum value of $m_M$, this mixing angle is allowed but will be tested by LHCb, ATLAS, and CMS \cite{Bolton:2019pcu}.
For 2 MeV $<m_{\nu_R}<2$ GeV existing data from CHARM, T2K, PIENU, and Borexino already rule this out \cite{CHARM:1985nku,CHARMII:1994jjr,T2K:2019jwa,Bryman:2019bjg,Borexino:2013bot}.
Sterile neutrinos with $m_{\nu_R}$ below 2 MeV are allowed.
The entire region below 1 GeV is in (model dependent) tension with cosmological results from BBN as well as combined CMB and BAO results \cite{Ruchayskiy:2012si,Vincent:2014rja}.

\textbf{Following the neutrino unitarity constraints}:
While the gallium result could be the first hint of the very large mixing this scenario predicts, we now continue as if it is disfavored, as suggested by solar neutrino data\footnote{See e.g.~\cite{Davoudiasl:2023uiq,Brdar:2023cms} for scenarios with a sterile neutrino compatible with the gallium measurements that also evades the solar neutrino constraints.}.
This constraint cannot be evaded by increasing the number of new neutrinos, e.g.~increasing $N$ in eq.~\ref{eq:tanphi}, either as eq.~\ref{eq:charge} for $N$ steriles with identical charge $B-L=1$  always reads
\begin{align}
    \varepsilon_\nu=-\varepsilon_n(1-2\delta_e)
\end{align}
independent of the number of steriles.

However one way to circumvent the unitarity violation constraints is to change the charge assignments of the new particles.
Since we need to cancel the neutrino charge which has $B-L=-1$, instead of adding in a new vectorlike neutrino with $B-L=1$, we assign the new neutrino charge $z_{\nu_4}>1$, which modifies eq.~\ref{eq:neutrino neutralization} to
\begin{equation}
\varepsilon_\nu=-\varepsilon_n(\cos^2\theta-z_{\nu_4}\sin^2\theta)\,.
\end{equation}
Since the new fermions are all vectorlike, the anomalies are automatically cancelled.
In this scenario we have that the new Higgs scalar has charge $z_{\nu_4}+1$.
We also see that, while this new Higgs scalar in the previous case automatically provided a Majorana mass term to the right handed neutrino mixing with the active neutrinos to give a traditional seesaw mass to those neutrinos, with the larger charge assignments $z_{\nu_4}>1$, this is no longer possible.
The active neutrinos can still get their masses from any number of scenarios including Dirac masses only, or via a seesaw with a third Higgs boson.

\begin{figure}
\centering
\includegraphics[width=\columnwidth]{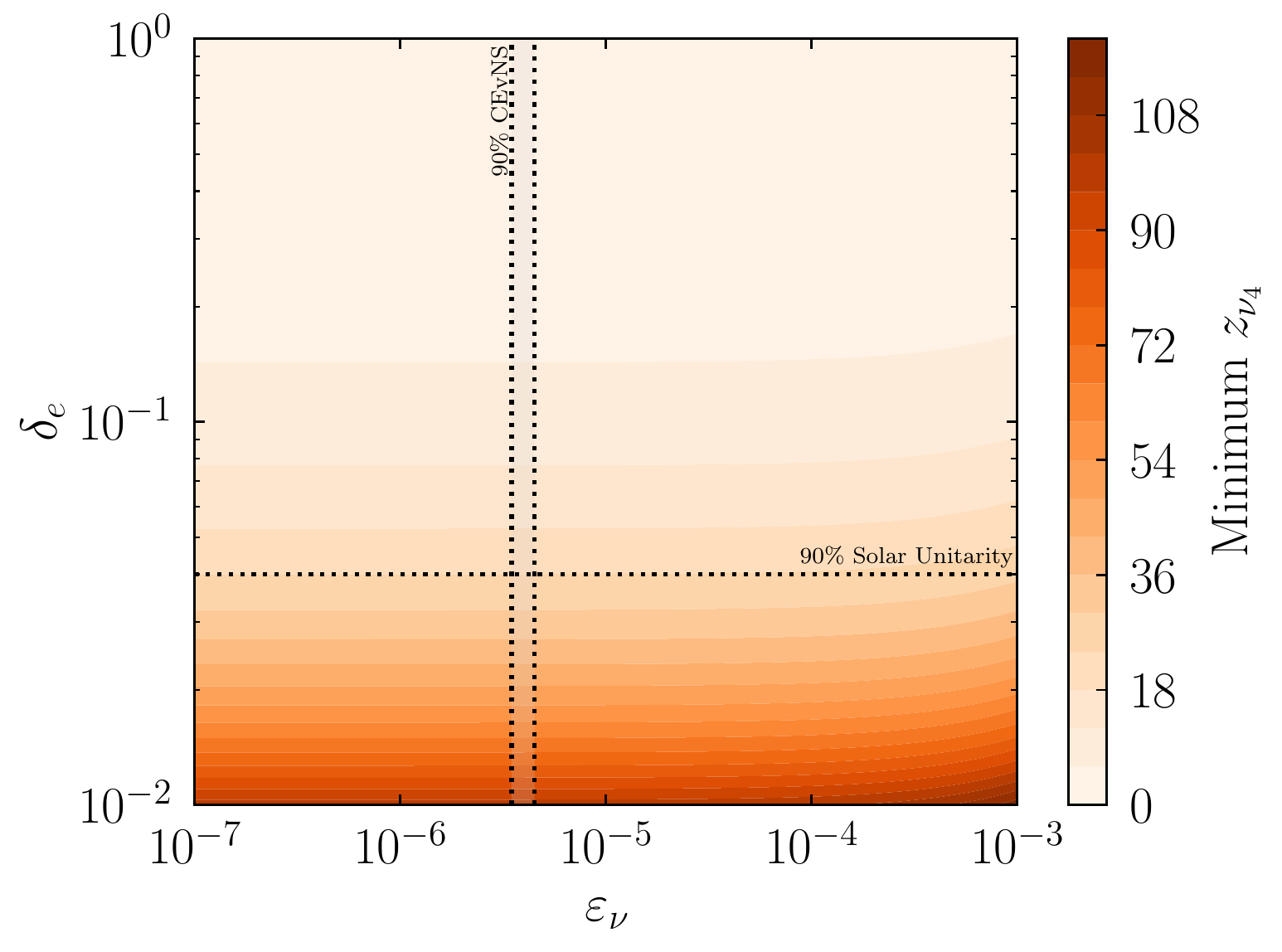}
\caption{The minimum required charge in the anomaly free $U(1)_{B-L}$ scenario on the new Dirac neutrino, $z_{\nu_4}$, to sufficiently neutralize the active neutrino charge below the limit from CEvNS (shaded region is due to the uncertainty in the Be isospin effects) shown as the vertical shaded region, while remaining consistent unitarity probes shown as the horizontal dotted line.
See eq.~\ref{eq:charge}.
The regions above and to the right of the dotted lines are disfavored.}
\label{fig:charge}
\end{figure}

Given a maximum deviation on the unitarity of the $\nu_e$ row of $\delta_e$, the charge of $\nu_4$ must be greater than
\begin{equation}
z_{\nu_4}\ge\frac{\frac{|\varepsilon_\nu|}{|\varepsilon_n|}+(1-\delta_e)}{\delta_e}\approx\frac1{\delta_e}\,,
\label{eq:charge}
\end{equation}
where the approximation applies when $|\varepsilon_\nu|\ll|\varepsilon_n|$ and $\delta_e\ll1$.
Thus we need $z_{\nu_4}\gtrsim24$ which allows one to evade the unitarity constraints on $\delta_e$ and neutralize the neutrino charge to below the CEvNS limit on $\varepsilon_\nu$.
The behavior of eq.~\ref{eq:charge} is shown in fig.~\ref{fig:charge}\footnote{The vertical CEvNS line assumes the no-isospin effects interpretation of the ATOMKI data, but changes to it clearly do not affect the minimum $z_{\nu_4}$ required.}.
The small mixing angle required by the unitarity constraint can be easily achieved by pushing $m_{\nu_4}$ up to $\sim135$ GeV in eq.~\ref{eq:tanphi} which is safe from constraints.

Future solar neutrino measurements from DUNE and HK as well as reactor measurements from JUNO will improve this unitarity constraint and either detect a deviation or further constrain $\nu_e$ row normalization unitarity thus increasing the required charge.
On the other hand, improvements to the constraint on $\varepsilon_\nu$ from e.g.~CEvNS will not increase the required charge since it is already known that $|\varepsilon_\nu|\ll|\varepsilon_n|$.

There may be additional ways to suppress the $\varepsilon_\nu$ mixing in a $U(1)_{B-L}$ scenario without the addition of new neutrinos, however these scenarios tend to be even more baroque. Alternatively, one could study a different gauge symmetry instead of $U(1)_{B-L}$ however also in this case the neutrino couplings need to be suppressed via the introduction of additional fermions \cite{Gu:2016ege} as a vector boson which couples directly to charged leptons automatically also couples to neutrinos.

\subsection{Small \texorpdfstring{$\varepsilon_e$}{varepsilon e} Scenarios}
One could attempt to set $|\varepsilon_e|$ much lower, below the limit from E137 \cite{Bjorken:1988as} and from supernova \cite{Kazanas:2014mca}.
This would automatically ensure that the neutrino bounds are evaded in e.g.~a $U(1)_{B-L}$ model, as typically the neutrino coupling is similar to the electron coupling, without relying on the specifics of any additional model building.
Such scenarios experience other problems, however.

If we consider the strongest constraints on $\varepsilon_e$ in this region, we find that the largest $\varepsilon_e$ can be is $10^{-12}$, at the limit from supernova \cite{Kazanas:2014mca}.
In this case $X$ would not decay in time for the ATOMKI experiments, which require that the dominant $\varepsilon$ contributing to its decay width to be $\gtrsim10^{-5}$.
Thus we must introduce a new dark fermion $\psi$ that couples to $X$ at $\varepsilon_\psi=10^{-5}$.
While this satisfies the lifetime constraint, a new problem arises.
Since
\begin{equation}
{\rm BR}(X\to e^+e^-)=\frac{\varepsilon_e^2}{\varepsilon_\psi^2}=10^{-14}\,,
\end{equation}
we must increase $\varepsilon_n$ and $\varepsilon_p$ by a factor of $10^7$ to get the correct widths to explain ATOMKI shown in fig.~\ref{fig:angular width}, at which point the couplings are well past the unitarity limit.

If the strong supernova constraints are ignored, as may be the case in the presence of additional new physics, at the limit from E137, $\varepsilon_e=10^{-7}$ \cite{Bjorken:1988as}.
In this case we would require increasing $\varepsilon_n$ and $\varepsilon_p$ by only a factor of $10^2$ which remains below the unitarity limit.
Nonetheless, $\varepsilon_n$ and $\varepsilon_p$ are now too large.
The constraint on $\varepsilon_p$ from $\pi^0$ decays is no longer relevant due to the $1/\sqrt{{\rm BR}(X\to e^+e^-)}$ factor and we require $\varepsilon_p\sim(0.1-0.2)$.
Then, we require $\varepsilon_n\sim(0.4-0.5)$ which is in considerable tension with the constraint from neutron-lead scattering \cite{Barbieri:1975xy,Feng:2016ysn} which is $\varepsilon_n<0.02$ for a mediator at 17 MeV.

\begin{center}
\pgfornament[width=0.618\columnwidth]{88}
\end{center}

To summarize this section we show that, while it is possible to realize a moderately ``upphobic" scenario by, for example gauging baryon number or $B-L$ and invoke kinetic mixing with the photon which allows different couplings of a mediator of neutrons and protons, these models face severe constraints largely from the neutrino sector and the fact that $X$ must decay within the detector.
In addition, we confirm previous results that small $\varepsilon_e$ scenarios cannot lead to viable models given existing constraints.
We find that one should consider one of the following three scenarios to achieve a viable model:
\begin{enumerate}
\item A flavor non-universal $U(1)_X$ model without the introduction of new fermions or an anomalous $U(1)_B$ scenario which requires additional quarks to cancel the anomalies.
\item A $U(1)_{B-L}$ scenario that explains neutrino masses with an additional heavy neutrino at 50 GeV $\lesssim m_{\nu_4}\lesssim(70-90)$ GeV and large mixing consistent with the gallium anomaly, but in tension with solar neutrinos. Additionally a Majorana neutrino with MeV-GeV mass is predicted which can be tested with upcoming experiments. 
\item A $U(1)_{B-L}$ scenario with an additional heavy neutrino at $m_{\nu_4}\gtrsim135$ GeV and large $B-L$ charges.
\end{enumerate}
Additional more involved models are likely possible as well, see e.g.~\cite{Bordes:2019qjk,Nam:2019osu}.

\section{Conclusions}
\label{sec:conclusion}
ATOMKI has reported several measurements that indicate new physics at high significance.
While their results have not been directly tested elsewhere, they are compelling due to their agreement in the implied mass of the particle from the measurements of the opening angles.
While it is unambiguous that they seem to point to a new particle with a mass just below 17 MeV, the nature of that particle is unclear, as well as any new dark sector it may provide a window into.

We provide an up-to-date statistical test of the data.
We include angular and width data from measurements of three separate targets and separately constrain the coupling to protons and neutrons as well as the new particle's mass.
We find that there are some non-trivial degeneracies.
We also find that, while the different measurements do not perfectly agree with each other, the internal tension ranges from moderate to significant depending on the assumptions on the nuclear physics treatment of Be, but is smaller than the large preference for new physics over the Standard Model.
Additionally, we find that in the scenario without isospin effects the best fit value of the mediator coupling to protons slightly exceeds the 90\% C.L.~bounds from $\pi^0$ decays.
If this anomaly is real, this could be an indication of the nature of the isospin corrections to Be or an indication that perhaps one of either the Be or C data sets has issues.

Reviewing other constraints on MeV scale physics makes it clear that the model building space is fairly constrained.
Notably the latest reactor neutrino measurements and the unitarity of the neutrino mixing matrix place key constraints.
We find that it is not possible to consistently explain the ATOMKI data with just one new particle and outline a set of relatively minimal scenarios in several different directions to generally illustrate the minimal model building requirements to explain the anomaly.
In addition, since the parameter space is somewhat tightly constrained, we anticipate that a confirmation could happen elsewhere soon, or the constraints will require even more complicated models to explain the ATOMKI data.
In fact, measurements of $g_e-2$ show a slight anomaly in the relevant region of parameter space.

In the future constraints from LHCb \cite{Ilten:2015hya}, DarkQuest \cite{Apyan:2022tsd}, FASER \cite{FASER:2018eoc}, NA64 \cite{NA64:2020xxh}, Mu3E phase II \cite{Echenard:2014lma}, BESIII \cite{Chen:2016dhm},  and experiments using rare pion or kaon decays \cite{Chiang:2016cyf}  will further test the couplings of $X$ to quarks and electrons, potentially even closing the entire allowed parameter space in a model independent fashion.
Also upcoming neutrino oscillation experiments as well as CEvNS experiments will improve the constraints on light mediators coupling to neutrinos and improving bounds on neutrino unitarity making it more and more challenging to develop self-consistent anomaly free models that explain the ATOMKI anomalies.

Furthermore, several experiments are planned to directly test the ATOMKI anomaly like DarkLight at the TRIUMF ARIEL e-linac \cite{Azuelos:2022nbu,DarkLight:2022uji}, a recently approved electron scattering experiment at Jefferson Lab \cite{Dutta:2023ifr} as well as the PADME experiment \cite{Darme:2022zfw,PADME:2022xly}, see \cite{Alves:2023ree} for a discussion of ongoing and upcoming efforts to test this anomaly. 

While the model building to explain ATOMKI is somewhat involved, given the relatively compelling nature of the anomalies we anticipate an interesting story will evolve in the coming years, regardless of the outcome.

\begin{acknowledgments}
We thank I.~Brivio, J.~Feng, and M.~Hostert for helpful comments.
PBD acknowledges support by the United States Department of Energy under Grant Contract No.~DE-SC0012704. JG thanks the HET group at BNL for kind hospitality during the writing of the paper.
\end{acknowledgments}

\appendix
\section{Further constraints on \texorpdfstring{$X$}{X}}
\label{sec:constraints}
In sec.~\ref{sec:con_dom} we collected the dominant constraints on $X$. Here we mention sub-dominant constraints which are nevertheless important for the validity of the model.

A constraint on the coupling of the $X$ boson to electrons comes from the required lifetime of $X$ in the ATOMKI experiment. Following \cite{Feng:2016ysn} we use that the distance between the target where the excited nuclear state is formed and the detector is $\mathcal{O}(\text{cm})$. We then require that $X$ propagates no more than 1 cm from its production point before it decays into electrons which leads to a constraint on its coupling to electrons as $\varepsilon_e>1.3\times 10^{-5}\times\sqrt{\text{BR}(X\to e^+e^-)}$.

NA64 conducted also a search for $X$ using its invisible decays in the process $e^-Z\to e^-Z X$, $X\to$ invisible  \cite{NA64:2021xzo}.
The constraint is $\varepsilon_e< (5.2\times10^{-5})/\sqrt{\text{BR}(X\to \text{inv})}$ for $m_X=17$ MeV at 90\% C.L.

A constraint from neutrino-electron scattering experiments bounds the product of $\varepsilon_e\varepsilon_{\nu_e}$.
The TEXONO experiment provides the strongest constraints for  $m_X\approx 17$ MeV \cite{TEXONO:2009knm} of $\sqrt{|\varepsilon_e\varepsilon_{\nu_e}|}<7\times 10^{-5}$ for $\varepsilon_e\varepsilon_{\nu_e}>0$,
$\sqrt{|\varepsilon_e\varepsilon_{\nu_e}|}<3\times 10^{-4}$ for $\varepsilon_e\varepsilon_{\nu_e}<0$ at 90\% C.L.

A weak constraint on $X$ coupling directly to protons exists by constraints on the temperature of the Sun which is determined by measuring the high energy $^8$B neutrino flux and combining with KamLAND measurements of the solar oscillation parameters \cite{Suliga:2020lir}.

These constraints are not dominant in the context of ATOMKI, but are orthogonal and depend on a different combination of parameters than the leading constraints.
As these constraints improve in the future, the dominant constraints may change in nontrivial ways.
Additional model-dependent constraints may also exist.

\bibliography{main}

\begin{thebibliography}{119}%
\makeatletter
\providecommand \@ifxundefined [1]{%
 \@ifx{#1\undefined}
}%
\providecommand \@ifnum [1]{%
 \ifnum #1\expandafter \@firstoftwo
 \else \expandafter \@secondoftwo
 \fi
}%
\providecommand \@ifx [1]{%
 \ifx #1\expandafter \@firstoftwo
 \else \expandafter \@secondoftwo
 \fi
}%
\providecommand \natexlab [1]{#1}%
\providecommand \enquote  [1]{``#1''}%
\providecommand \bibnamefont  [1]{#1}%
\providecommand \bibfnamefont [1]{#1}%
\providecommand \citenamefont [1]{#1}%
\providecommand \href@noop [0]{\@secondoftwo}%
\providecommand \href [0]{\begingroup \@sanitize@url \@href}%
\providecommand \@href[1]{\@@startlink{#1}\@@href}%
\providecommand \@@href[1]{\endgroup#1\@@endlink}%
\providecommand \@sanitize@url [0]{\catcode `\\12\catcode `\$12\catcode
  `\&12\catcode `\#12\catcode `\^12\catcode `\_12\catcode `\%12\relax}%
\providecommand \@@startlink[1]{}%
\providecommand \@@endlink[0]{}%
\providecommand \url  [0]{\begingroup\@sanitize@url \@url }%
\providecommand \@url [1]{\endgroup\@href {#1}{\urlprefix }}%
\providecommand \urlprefix  [0]{URL }%
\providecommand \Eprint [0]{\href }%
\providecommand \doibase [0]{http://dx.doi.org/}%
\providecommand \selectlanguage [0]{\@gobble}%
\providecommand \bibinfo  [0]{\@secondoftwo}%
\providecommand \bibfield  [0]{\@secondoftwo}%
\providecommand \translation [1]{[#1]}%
\providecommand \BibitemOpen [0]{}%
\providecommand \bibitemStop [0]{}%
\providecommand \bibitemNoStop [0]{.\EOS\space}%
\providecommand \EOS [0]{\spacefactor3000\relax}%
\providecommand \BibitemShut  [1]{\csname bibitem#1\endcsname}%
\let\auto@bib@innerbib\@empty
%</preamble>
\bibitem [{\citenamefont {Krasznahorkay}\ \emph {et~al.}(2016)\citenamefont
  {Krasznahorkay} \emph {et~al.}}]{Krasznahorkay:2015iga}%
  \BibitemOpen
  \bibfield  {author} {\bibinfo {author} {\bibfnamefont {A.~J.}\ \bibnamefont
  {Krasznahorkay}} \emph {et~al.},\ }\href {\doibase
  10.1103/PhysRevLett.116.042501} {\bibfield  {journal} {\bibinfo  {journal}
  {Phys. Rev. Lett.}\ }\textbf {\bibinfo {volume} {116}},\ \bibinfo {pages}
  {042501} (\bibinfo {year} {2016})},\ \Eprint
  {http://arxiv.org/abs/1504.01527} {arXiv:1504.01527 [nucl-ex]} \BibitemShut
  {NoStop}%
\bibitem [{\citenamefont {Krasznahorkay}\ \emph {et~al.}(2019)\citenamefont
  {Krasznahorkay} \emph {et~al.}}]{Krasznahorkay:2019lyl}%
  \BibitemOpen
  \bibfield  {author} {\bibinfo {author} {\bibfnamefont {A.~J.}\ \bibnamefont
  {Krasznahorkay}} \emph {et~al.},\ }\href@noop {} {\  (\bibinfo {year}
  {2019})},\ \Eprint {http://arxiv.org/abs/1910.10459} {arXiv:1910.10459
  [nucl-ex]} \BibitemShut {NoStop}%
\bibitem [{\citenamefont {Krasznahorkay}\ \emph {et~al.}(2021)\citenamefont
  {Krasznahorkay}, \citenamefont {Csatl\'os}, \citenamefont {Csige},
  \citenamefont {Guly\'as}, \citenamefont {Krasznahorkay}, \citenamefont
  {Nyak\'o}, \citenamefont {Rajta}, \citenamefont {Tim\'ar}, \citenamefont
  {Vajda},\ and\ \citenamefont {Sas}}]{Krasznahorkay:2021joi}%
  \BibitemOpen
  \bibfield  {author} {\bibinfo {author} {\bibfnamefont {A.~J.}\ \bibnamefont
  {Krasznahorkay}}, \bibinfo {author} {\bibfnamefont {M.}~\bibnamefont
  {Csatl\'os}}, \bibinfo {author} {\bibfnamefont {L.}~\bibnamefont {Csige}},
  \bibinfo {author} {\bibfnamefont {J.}~\bibnamefont {Guly\'as}}, \bibinfo
  {author} {\bibfnamefont {A.}~\bibnamefont {Krasznahorkay}}, \bibinfo {author}
  {\bibfnamefont {B.~M.}\ \bibnamefont {Nyak\'o}}, \bibinfo {author}
  {\bibfnamefont {I.}~\bibnamefont {Rajta}}, \bibinfo {author} {\bibfnamefont
  {J.}~\bibnamefont {Tim\'ar}}, \bibinfo {author} {\bibfnamefont
  {I.}~\bibnamefont {Vajda}}, \ and\ \bibinfo {author} {\bibfnamefont {N.~J.}\
  \bibnamefont {Sas}},\ }\href {\doibase 10.1103/PhysRevC.104.044003}
  {\bibfield  {journal} {\bibinfo  {journal} {Phys. Rev. C}\ }\textbf {\bibinfo
  {volume} {104}},\ \bibinfo {pages} {044003} (\bibinfo {year} {2021})},\
  \Eprint {http://arxiv.org/abs/2104.10075} {arXiv:2104.10075 [nucl-ex]}
  \BibitemShut {NoStop}%
\bibitem [{\citenamefont {Krasznahorkay}\ \emph {et~al.}(2022)\citenamefont
  {Krasznahorkay} \emph {et~al.}}]{Krasznahorkay:2022pxs}%
  \BibitemOpen
  \bibfield  {author} {\bibinfo {author} {\bibfnamefont {A.~J.}\ \bibnamefont
  {Krasznahorkay}} \emph {et~al.},\ }\href {\doibase
  10.1103/PhysRevC.106.L061601} {\bibfield  {journal} {\bibinfo  {journal}
  {Phys. Rev. C}\ }\textbf {\bibinfo {volume} {106}},\ \bibinfo {pages}
  {L061601} (\bibinfo {year} {2022})},\ \Eprint
  {http://arxiv.org/abs/2209.10795} {arXiv:2209.10795 [nucl-ex]} \BibitemShut
  {NoStop}%
\bibitem [{\citenamefont {Alves}\ \emph {et~al.}(2023)\citenamefont {Alves}
  \emph {et~al.}}]{Alves:2023ree}%
  \BibitemOpen
  \bibfield  {author} {\bibinfo {author} {\bibfnamefont {D.~S.~M.}\
  \bibnamefont {Alves}} \emph {et~al.},\ }\href {\doibase
  10.1140/epjc/s10052-023-11271-x} {\bibfield  {journal} {\bibinfo  {journal}
  {Eur. Phys. J. C}\ }\textbf {\bibinfo {volume} {83}},\ \bibinfo {pages} {230}
  (\bibinfo {year} {2023})}\BibitemShut {NoStop}%
\bibitem [{\citenamefont {Zhang}\ and\ \citenamefont
  {Miller}(2017)}]{Zhang:2017zap}%
  \BibitemOpen
  \bibfield  {author} {\bibinfo {author} {\bibfnamefont {X.}~\bibnamefont
  {Zhang}}\ and\ \bibinfo {author} {\bibfnamefont {G.~A.}\ \bibnamefont
  {Miller}},\ }\href {\doibase 10.1016/j.physletb.2017.08.013} {\bibfield
  {journal} {\bibinfo  {journal} {Phys. Lett. B}\ }\textbf {\bibinfo {volume}
  {773}},\ \bibinfo {pages} {159} (\bibinfo {year} {2017})},\ \Eprint
  {http://arxiv.org/abs/1703.04588} {arXiv:1703.04588 [nucl-th]} \BibitemShut
  {NoStop}%
\bibitem [{\citenamefont {Koch}(2021)}]{Koch:2020ouk}%
  \BibitemOpen
  \bibfield  {author} {\bibinfo {author} {\bibfnamefont {B.}~\bibnamefont
  {Koch}},\ }\href {\doibase 10.1016/j.nuclphysa.2021.122143} {\bibfield
  {journal} {\bibinfo  {journal} {Nucl. Phys. A}\ }\textbf {\bibinfo {volume}
  {1008}},\ \bibinfo {pages} {122143} (\bibinfo {year} {2021})},\ \Eprint
  {http://arxiv.org/abs/2003.05722} {arXiv:2003.05722 [hep-ph]} \BibitemShut
  {NoStop}%
\bibitem [{\citenamefont {Chen}(2020)}]{Chen:2020arr}%
  \BibitemOpen
  \bibfield  {author} {\bibinfo {author} {\bibfnamefont {H.-X.}\ \bibnamefont
  {Chen}},\ }\href@noop {} {\  (\bibinfo {year} {2020})},\ \Eprint
  {http://arxiv.org/abs/2006.01018} {arXiv:2006.01018 [hep-ph]} \BibitemShut
  {NoStop}%
\bibitem [{\citenamefont {Kubarovsky}\ \emph {et~al.}(2022)\citenamefont
  {Kubarovsky}, \citenamefont {West},\ and\ \citenamefont
  {Brodsky}}]{Kubarovsky:2022zxm}%
  \BibitemOpen
  \bibfield  {author} {\bibinfo {author} {\bibfnamefont {V.}~\bibnamefont
  {Kubarovsky}}, \bibinfo {author} {\bibfnamefont {J.~R.}\ \bibnamefont
  {West}}, \ and\ \bibinfo {author} {\bibfnamefont {S.~J.}\ \bibnamefont
  {Brodsky}},\ }\href@noop {} {\  (\bibinfo {year} {2022})},\ \Eprint
  {http://arxiv.org/abs/2206.14441} {arXiv:2206.14441 [hep-ph]} \BibitemShut
  {NoStop}%
\bibitem [{\citenamefont {Wong}(2022)}]{Wong:2022kyg}%
  \BibitemOpen
  \bibfield  {author} {\bibinfo {author} {\bibfnamefont {C.-Y.}\ \bibnamefont
  {Wong}},\ }in\ \href@noop {} {\emph {\bibinfo {booktitle} {{Shedding light on
  X17}}}}\ (\bibinfo {year} {2022})\ \Eprint {http://arxiv.org/abs/2201.09764}
  {arXiv:2201.09764 [hep-ph]} \BibitemShut {NoStop}%
\bibitem [{\citenamefont {Aleksejevs}\ \emph {et~al.}(2021)\citenamefont
  {Aleksejevs}, \citenamefont {Barkanova}, \citenamefont {Kolomensky},\ and\
  \citenamefont {Sheff}}]{Aleksejevs:2021zjw}%
  \BibitemOpen
  \bibfield  {author} {\bibinfo {author} {\bibfnamefont {A.}~\bibnamefont
  {Aleksejevs}}, \bibinfo {author} {\bibfnamefont {S.}~\bibnamefont
  {Barkanova}}, \bibinfo {author} {\bibfnamefont {Y.~G.}\ \bibnamefont
  {Kolomensky}}, \ and\ \bibinfo {author} {\bibfnamefont {B.}~\bibnamefont
  {Sheff}},\ }\href@noop {} {\  (\bibinfo {year} {2021})},\ \Eprint
  {http://arxiv.org/abs/2102.01127} {arXiv:2102.01127 [hep-ph]} \BibitemShut
  {NoStop}%
\bibitem [{\citenamefont {Hayes}\ \emph {et~al.}(2022)\citenamefont {Hayes},
  \citenamefont {Friar}, \citenamefont {Hale},\ and\ \citenamefont
  {Garvey}}]{Hayes:2021hin}%
  \BibitemOpen
  \bibfield  {author} {\bibinfo {author} {\bibfnamefont {A.~C.}\ \bibnamefont
  {Hayes}}, \bibinfo {author} {\bibfnamefont {J.~L.}\ \bibnamefont {Friar}},
  \bibinfo {author} {\bibfnamefont {G.~M.}\ \bibnamefont {Hale}}, \ and\
  \bibinfo {author} {\bibfnamefont {G.~T.}\ \bibnamefont {Garvey}},\ }\href
  {\doibase 10.1103/PhysRevC.105.055502} {\bibfield  {journal} {\bibinfo
  {journal} {Phys. Rev. C}\ }\textbf {\bibinfo {volume} {105}},\ \bibinfo
  {pages} {055502} (\bibinfo {year} {2022})},\ \Eprint
  {http://arxiv.org/abs/2106.06834} {arXiv:2106.06834 [nucl-th]} \BibitemShut
  {NoStop}%
\bibitem [{\citenamefont {Viviani}\ \emph {et~al.}(2022)\citenamefont
  {Viviani}, \citenamefont {Filandri}, \citenamefont {Girlanda}, \citenamefont
  {Gustavino}, \citenamefont {Kievsky}, \citenamefont {Marcucci},\ and\
  \citenamefont {Schiavilla}}]{Viviani:2021stx}%
  \BibitemOpen
  \bibfield  {author} {\bibinfo {author} {\bibfnamefont {M.}~\bibnamefont
  {Viviani}}, \bibinfo {author} {\bibfnamefont {E.}~\bibnamefont {Filandri}},
  \bibinfo {author} {\bibfnamefont {L.}~\bibnamefont {Girlanda}}, \bibinfo
  {author} {\bibfnamefont {C.}~\bibnamefont {Gustavino}}, \bibinfo {author}
  {\bibfnamefont {A.}~\bibnamefont {Kievsky}}, \bibinfo {author} {\bibfnamefont
  {L.~E.}\ \bibnamefont {Marcucci}}, \ and\ \bibinfo {author} {\bibfnamefont
  {R.}~\bibnamefont {Schiavilla}},\ }\href {\doibase
  10.1103/PhysRevC.105.014001} {\bibfield  {journal} {\bibinfo  {journal}
  {Phys. Rev. C}\ }\textbf {\bibinfo {volume} {105}},\ \bibinfo {pages}
  {014001} (\bibinfo {year} {2022})},\ \Eprint
  {http://arxiv.org/abs/2104.07808} {arXiv:2104.07808 [nucl-th]} \BibitemShut
  {NoStop}%
\bibitem [{\citenamefont {Feng}\ \emph {et~al.}(2017)\citenamefont {Feng},
  \citenamefont {Fornal}, \citenamefont {Galon}, \citenamefont {Gardner},
  \citenamefont {Smolinsky}, \citenamefont {Tait},\ and\ \citenamefont
  {Tanedo}}]{Feng:2016ysn}%
  \BibitemOpen
  \bibfield  {author} {\bibinfo {author} {\bibfnamefont {J.~L.}\ \bibnamefont
  {Feng}}, \bibinfo {author} {\bibfnamefont {B.}~\bibnamefont {Fornal}},
  \bibinfo {author} {\bibfnamefont {I.}~\bibnamefont {Galon}}, \bibinfo
  {author} {\bibfnamefont {S.}~\bibnamefont {Gardner}}, \bibinfo {author}
  {\bibfnamefont {J.}~\bibnamefont {Smolinsky}}, \bibinfo {author}
  {\bibfnamefont {T.~M.~P.}\ \bibnamefont {Tait}}, \ and\ \bibinfo {author}
  {\bibfnamefont {P.}~\bibnamefont {Tanedo}},\ }\href {\doibase
  10.1103/PhysRevD.95.035017} {\bibfield  {journal} {\bibinfo  {journal} {Phys.
  Rev. D}\ }\textbf {\bibinfo {volume} {95}},\ \bibinfo {pages} {035017}
  (\bibinfo {year} {2017})},\ \Eprint {http://arxiv.org/abs/1608.03591}
  {arXiv:1608.03591 [hep-ph]} \BibitemShut {NoStop}%
\bibitem [{\citenamefont {Feng}\ \emph {et~al.}(2016)\citenamefont {Feng},
  \citenamefont {Fornal}, \citenamefont {Galon}, \citenamefont {Gardner},
  \citenamefont {Smolinsky}, \citenamefont {Tait},\ and\ \citenamefont
  {Tanedo}}]{Feng:2016jff}%
  \BibitemOpen
  \bibfield  {author} {\bibinfo {author} {\bibfnamefont {J.~L.}\ \bibnamefont
  {Feng}}, \bibinfo {author} {\bibfnamefont {B.}~\bibnamefont {Fornal}},
  \bibinfo {author} {\bibfnamefont {I.}~\bibnamefont {Galon}}, \bibinfo
  {author} {\bibfnamefont {S.}~\bibnamefont {Gardner}}, \bibinfo {author}
  {\bibfnamefont {J.}~\bibnamefont {Smolinsky}}, \bibinfo {author}
  {\bibfnamefont {T.~M.~P.}\ \bibnamefont {Tait}}, \ and\ \bibinfo {author}
  {\bibfnamefont {P.}~\bibnamefont {Tanedo}},\ }\href {\doibase
  10.1103/PhysRevLett.117.071803} {\bibfield  {journal} {\bibinfo  {journal}
  {Phys. Rev. Lett.}\ }\textbf {\bibinfo {volume} {117}},\ \bibinfo {pages}
  {071803} (\bibinfo {year} {2016})},\ \Eprint
  {http://arxiv.org/abs/1604.07411} {arXiv:1604.07411 [hep-ph]} \BibitemShut
  {NoStop}%
\bibitem [{\citenamefont {Puli\c{c}e}(2021)}]{Pulice:2019xel}%
  \BibitemOpen
  \bibfield  {author} {\bibinfo {author} {\bibfnamefont {B.}~\bibnamefont
  {Puli\c{c}e}},\ }\href {\doibase 10.1016/j.cjph.2020.11.021} {\bibfield
  {journal} {\bibinfo  {journal} {Chin. J. Phys.}\ }\textbf {\bibinfo {volume}
  {71}},\ \bibinfo {pages} {506} (\bibinfo {year} {2021})},\ \Eprint
  {http://arxiv.org/abs/1911.10482} {arXiv:1911.10482 [hep-ph]} \BibitemShut
  {NoStop}%
\bibitem [{\citenamefont {Feng}\ \emph {et~al.}(2020)\citenamefont {Feng},
  \citenamefont {Tait},\ and\ \citenamefont {Verhaaren}}]{Feng:2020mbt}%
  \BibitemOpen
  \bibfield  {author} {\bibinfo {author} {\bibfnamefont {J.~L.}\ \bibnamefont
  {Feng}}, \bibinfo {author} {\bibfnamefont {T.~M.~P.}\ \bibnamefont {Tait}}, \
  and\ \bibinfo {author} {\bibfnamefont {C.~B.}\ \bibnamefont {Verhaaren}},\
  }\href {\doibase 10.1103/PhysRevD.102.036016} {\bibfield  {journal} {\bibinfo
   {journal} {Phys. Rev. D}\ }\textbf {\bibinfo {volume} {102}},\ \bibinfo
  {pages} {036016} (\bibinfo {year} {2020})},\ \Eprint
  {http://arxiv.org/abs/2006.01151} {arXiv:2006.01151 [hep-ph]} \BibitemShut
  {NoStop}%
\bibitem [{\citenamefont {Nomura}\ and\ \citenamefont
  {Sanyal}(2021)}]{Nomura:2020kcw}%
  \BibitemOpen
  \bibfield  {author} {\bibinfo {author} {\bibfnamefont {T.}~\bibnamefont
  {Nomura}}\ and\ \bibinfo {author} {\bibfnamefont {P.}~\bibnamefont
  {Sanyal}},\ }\href {\doibase 10.1007/JHEP05(2021)232} {\bibfield  {journal}
  {\bibinfo  {journal} {JHEP}\ }\textbf {\bibinfo {volume} {05}},\ \bibinfo
  {pages} {232} (\bibinfo {year} {2021})},\ \Eprint
  {http://arxiv.org/abs/2010.04266} {arXiv:2010.04266 [hep-ph]} \BibitemShut
  {NoStop}%
\bibitem [{\citenamefont {Kozaczuk}\ \emph {et~al.}(2017)\citenamefont
  {Kozaczuk}, \citenamefont {Morrissey},\ and\ \citenamefont
  {Stroberg}}]{Kozaczuk:2016nma}%
  \BibitemOpen
  \bibfield  {author} {\bibinfo {author} {\bibfnamefont {J.}~\bibnamefont
  {Kozaczuk}}, \bibinfo {author} {\bibfnamefont {D.~E.}\ \bibnamefont
  {Morrissey}}, \ and\ \bibinfo {author} {\bibfnamefont {S.~R.}\ \bibnamefont
  {Stroberg}},\ }\href {\doibase 10.1103/PhysRevD.95.115024} {\bibfield
  {journal} {\bibinfo  {journal} {Phys. Rev. D}\ }\textbf {\bibinfo {volume}
  {95}},\ \bibinfo {pages} {115024} (\bibinfo {year} {2017})},\ \Eprint
  {http://arxiv.org/abs/1612.01525} {arXiv:1612.01525 [hep-ph]} \BibitemShut
  {NoStop}%
\bibitem [{\citenamefont {Delle~Rose}\ \emph {et~al.}(2019)\citenamefont
  {Delle~Rose}, \citenamefont {Khalil}, \citenamefont {King},\ and\
  \citenamefont {Moretti}}]{DelleRose:2018pgm}%
  \BibitemOpen
  \bibfield  {author} {\bibinfo {author} {\bibfnamefont {L.}~\bibnamefont
  {Delle~Rose}}, \bibinfo {author} {\bibfnamefont {S.}~\bibnamefont {Khalil}},
  \bibinfo {author} {\bibfnamefont {S.~J.~D.}\ \bibnamefont {King}}, \ and\
  \bibinfo {author} {\bibfnamefont {S.}~\bibnamefont {Moretti}},\ }\href
  {\doibase 10.3389/fphy.2019.00073} {\bibfield  {journal} {\bibinfo  {journal}
  {Front. in Phys.}\ }\textbf {\bibinfo {volume} {7}},\ \bibinfo {pages} {73}
  (\bibinfo {year} {2019})},\ \Eprint {http://arxiv.org/abs/1812.05497}
  {arXiv:1812.05497 [hep-ph]} \BibitemShut {NoStop}%
\bibitem [{\citenamefont {Seto}\ and\ \citenamefont
  {Shimomura}(2021)}]{Seto:2020jal}%
  \BibitemOpen
  \bibfield  {author} {\bibinfo {author} {\bibfnamefont {O.}~\bibnamefont
  {Seto}}\ and\ \bibinfo {author} {\bibfnamefont {T.}~\bibnamefont
  {Shimomura}},\ }\href {\doibase 10.1007/JHEP04(2021)025} {\bibfield
  {journal} {\bibinfo  {journal} {JHEP}\ }\textbf {\bibinfo {volume} {04}},\
  \bibinfo {pages} {025} (\bibinfo {year} {2021})},\ \Eprint
  {http://arxiv.org/abs/2006.05497} {arXiv:2006.05497 [hep-ph]} \BibitemShut
  {NoStop}%
\bibitem [{\citenamefont {Barducci}\ and\ \citenamefont
  {Toni}(2023)}]{Barducci:2022lqd}%
  \BibitemOpen
  \bibfield  {author} {\bibinfo {author} {\bibfnamefont {D.}~\bibnamefont
  {Barducci}}\ and\ \bibinfo {author} {\bibfnamefont {C.}~\bibnamefont
  {Toni}},\ }\href {\doibase 10.1007/JHEP02(2023)154} {\bibfield  {journal}
  {\bibinfo  {journal} {JHEP}\ }\textbf {\bibinfo {volume} {02}},\ \bibinfo
  {pages} {154} (\bibinfo {year} {2023})},\ \Eprint
  {http://arxiv.org/abs/2212.06453} {arXiv:2212.06453 [hep-ph]} \BibitemShut
  {NoStop}%
\bibitem [{\citenamefont {Zhang}\ and\ \citenamefont
  {Miller}(2021)}]{Zhang:2020ukq}%
  \BibitemOpen
  \bibfield  {author} {\bibinfo {author} {\bibfnamefont {X.}~\bibnamefont
  {Zhang}}\ and\ \bibinfo {author} {\bibfnamefont {G.~A.}\ \bibnamefont
  {Miller}},\ }\href {\doibase 10.1016/j.physletb.2021.136061} {\bibfield
  {journal} {\bibinfo  {journal} {Phys. Lett. B}\ }\textbf {\bibinfo {volume}
  {813}},\ \bibinfo {pages} {136061} (\bibinfo {year} {2021})},\ \Eprint
  {http://arxiv.org/abs/2008.11288} {arXiv:2008.11288 [hep-ph]} \BibitemShut
  {NoStop}%
\bibitem [{\citenamefont {Sas}\ \emph {et~al.}(2022)\citenamefont {Sas} \emph
  {et~al.}}]{Sas:2022pgm}%
  \BibitemOpen
  \bibfield  {author} {\bibinfo {author} {\bibfnamefont {N.~J.}\ \bibnamefont
  {Sas}} \emph {et~al.},\ }\href@noop {} {\  (\bibinfo {year} {2022})},\
  \Eprint {http://arxiv.org/abs/2205.07744} {arXiv:2205.07744 [nucl-ex]}
  \BibitemShut {NoStop}%
\bibitem [{\citenamefont {Freedman}(1974)}]{Freedman:1973yd}%
  \BibitemOpen
  \bibfield  {author} {\bibinfo {author} {\bibfnamefont {D.~Z.}\ \bibnamefont
  {Freedman}},\ }\href {\doibase 10.1103/PhysRevD.9.1389} {\bibfield  {journal}
  {\bibinfo  {journal} {Phys. Rev. D}\ }\textbf {\bibinfo {volume} {9}},\
  \bibinfo {pages} {1389} (\bibinfo {year} {1974})}\BibitemShut {NoStop}%
\bibitem [{\citenamefont {Coloma}\ \emph
  {et~al.}(2017{\natexlab{a}})\citenamefont {Coloma}, \citenamefont {Denton},
  \citenamefont {Gonzalez-Garcia}, \citenamefont {Maltoni},\ and\ \citenamefont
  {Schwetz}}]{Coloma:2017egw}%
  \BibitemOpen
  \bibfield  {author} {\bibinfo {author} {\bibfnamefont {P.}~\bibnamefont
  {Coloma}}, \bibinfo {author} {\bibfnamefont {P.~B.}\ \bibnamefont {Denton}},
  \bibinfo {author} {\bibfnamefont {M.~C.}\ \bibnamefont {Gonzalez-Garcia}},
  \bibinfo {author} {\bibfnamefont {M.}~\bibnamefont {Maltoni}}, \ and\
  \bibinfo {author} {\bibfnamefont {T.}~\bibnamefont {Schwetz}},\ }\href
  {\doibase 10.1007/JHEP04(2017)116} {\bibfield  {journal} {\bibinfo  {journal}
  {JHEP}\ }\textbf {\bibinfo {volume} {04}},\ \bibinfo {pages} {116} (\bibinfo
  {year} {2017}{\natexlab{a}})},\ \Eprint {http://arxiv.org/abs/1701.04828}
  {arXiv:1701.04828 [hep-ph]} \BibitemShut {NoStop}%
\bibitem [{\citenamefont {Liao}\ and\ \citenamefont
  {Marfatia}(2017)}]{Liao:2017uzy}%
  \BibitemOpen
  \bibfield  {author} {\bibinfo {author} {\bibfnamefont {J.}~\bibnamefont
  {Liao}}\ and\ \bibinfo {author} {\bibfnamefont {D.}~\bibnamefont
  {Marfatia}},\ }\href {\doibase 10.1016/j.physletb.2017.10.046} {\bibfield
  {journal} {\bibinfo  {journal} {Phys. Lett. B}\ }\textbf {\bibinfo {volume}
  {775}},\ \bibinfo {pages} {54} (\bibinfo {year} {2017})},\ \Eprint
  {http://arxiv.org/abs/1708.04255} {arXiv:1708.04255 [hep-ph]} \BibitemShut
  {NoStop}%
\bibitem [{\citenamefont {Coloma}\ \emph
  {et~al.}(2017{\natexlab{b}})\citenamefont {Coloma}, \citenamefont
  {Gonzalez-Garcia}, \citenamefont {Maltoni},\ and\ \citenamefont
  {Schwetz}}]{Coloma:2017ncl}%
  \BibitemOpen
  \bibfield  {author} {\bibinfo {author} {\bibfnamefont {P.}~\bibnamefont
  {Coloma}}, \bibinfo {author} {\bibfnamefont {M.~C.}\ \bibnamefont
  {Gonzalez-Garcia}}, \bibinfo {author} {\bibfnamefont {M.}~\bibnamefont
  {Maltoni}}, \ and\ \bibinfo {author} {\bibfnamefont {T.}~\bibnamefont
  {Schwetz}},\ }\href {\doibase 10.1103/PhysRevD.96.115007} {\bibfield
  {journal} {\bibinfo  {journal} {Phys. Rev. D}\ }\textbf {\bibinfo {volume}
  {96}},\ \bibinfo {pages} {115007} (\bibinfo {year} {2017}{\natexlab{b}})},\
  \Eprint {http://arxiv.org/abs/1708.02899} {arXiv:1708.02899 [hep-ph]}
  \BibitemShut {NoStop}%
\bibitem [{\citenamefont {Denton}\ \emph {et~al.}(2018)\citenamefont {Denton},
  \citenamefont {Farzan},\ and\ \citenamefont {Shoemaker}}]{Denton:2018xmq}%
  \BibitemOpen
  \bibfield  {author} {\bibinfo {author} {\bibfnamefont {P.~B.}\ \bibnamefont
  {Denton}}, \bibinfo {author} {\bibfnamefont {Y.}~\bibnamefont {Farzan}}, \
  and\ \bibinfo {author} {\bibfnamefont {I.~M.}\ \bibnamefont {Shoemaker}},\
  }\href {\doibase 10.1007/JHEP07(2018)037} {\bibfield  {journal} {\bibinfo
  {journal} {JHEP}\ }\textbf {\bibinfo {volume} {07}},\ \bibinfo {pages} {037}
  (\bibinfo {year} {2018})},\ \Eprint {http://arxiv.org/abs/1804.03660}
  {arXiv:1804.03660 [hep-ph]} \BibitemShut {NoStop}%
\bibitem [{\citenamefont {Denton}\ and\ \citenamefont
  {Gehrlein}(2021)}]{Denton:2020hop}%
  \BibitemOpen
  \bibfield  {author} {\bibinfo {author} {\bibfnamefont {P.~B.}\ \bibnamefont
  {Denton}}\ and\ \bibinfo {author} {\bibfnamefont {J.}~\bibnamefont
  {Gehrlein}},\ }\href {\doibase 10.1007/JHEP04(2021)266} {\bibfield  {journal}
  {\bibinfo  {journal} {JHEP}\ }\textbf {\bibinfo {volume} {04}},\ \bibinfo
  {pages} {266} (\bibinfo {year} {2021})},\ \Eprint
  {http://arxiv.org/abs/2008.06062} {arXiv:2008.06062 [hep-ph]} \BibitemShut
  {NoStop}%
\bibitem [{\citenamefont {Aristizabal~Sierra}\ \emph
  {et~al.}(2022)\citenamefont {Aristizabal~Sierra}, \citenamefont {De~Romeri},\
  and\ \citenamefont {Papoulias}}]{AristizabalSierra:2022axl}%
  \BibitemOpen
  \bibfield  {author} {\bibinfo {author} {\bibfnamefont {D.}~\bibnamefont
  {Aristizabal~Sierra}}, \bibinfo {author} {\bibfnamefont {V.}~\bibnamefont
  {De~Romeri}}, \ and\ \bibinfo {author} {\bibfnamefont {D.~K.}\ \bibnamefont
  {Papoulias}},\ }\href {\doibase 10.1007/JHEP09(2022)076} {\bibfield
  {journal} {\bibinfo  {journal} {JHEP}\ }\textbf {\bibinfo {volume} {09}},\
  \bibinfo {pages} {076} (\bibinfo {year} {2022})},\ \Eprint
  {http://arxiv.org/abs/2203.02414} {arXiv:2203.02414 [hep-ph]} \BibitemShut
  {NoStop}%
\bibitem [{\citenamefont {Liao}\ \emph {et~al.}(2022)\citenamefont {Liao},
  \citenamefont {Liu},\ and\ \citenamefont {Marfatia}}]{Liao:2022hno}%
  \BibitemOpen
  \bibfield  {author} {\bibinfo {author} {\bibfnamefont {J.}~\bibnamefont
  {Liao}}, \bibinfo {author} {\bibfnamefont {H.}~\bibnamefont {Liu}}, \ and\
  \bibinfo {author} {\bibfnamefont {D.}~\bibnamefont {Marfatia}},\ }\href
  {\doibase 10.1103/PhysRevD.106.L031702} {\bibfield  {journal} {\bibinfo
  {journal} {Phys. Rev. D}\ }\textbf {\bibinfo {volume} {106}},\ \bibinfo
  {pages} {L031702} (\bibinfo {year} {2022})},\ \Eprint
  {http://arxiv.org/abs/2202.10622} {arXiv:2202.10622 [hep-ph]} \BibitemShut
  {NoStop}%
\bibitem [{\citenamefont {Denton}\ and\ \citenamefont
  {Gehrlein}(2022)}]{Denton:2022nol}%
  \BibitemOpen
  \bibfield  {author} {\bibinfo {author} {\bibfnamefont {P.~B.}\ \bibnamefont
  {Denton}}\ and\ \bibinfo {author} {\bibfnamefont {J.}~\bibnamefont
  {Gehrlein}},\ }\href {\doibase 10.1103/PhysRevD.106.015022} {\bibfield
  {journal} {\bibinfo  {journal} {Phys. Rev. D}\ }\textbf {\bibinfo {volume}
  {106}},\ \bibinfo {pages} {015022} (\bibinfo {year} {2022})},\ \Eprint
  {http://arxiv.org/abs/2204.09060} {arXiv:2204.09060 [hep-ph]} \BibitemShut
  {NoStop}%
\bibitem [{\citenamefont {Atzori~Corona}\ \emph {et~al.}(2022)\citenamefont
  {Atzori~Corona}, \citenamefont {Cadeddu}, \citenamefont {Cargioli},
  \citenamefont {Dordei}, \citenamefont {Giunti}, \citenamefont {Li},
  \citenamefont {Picciau}, \citenamefont {Ternes},\ and\ \citenamefont
  {Zhang}}]{AtzoriCorona:2022moj}%
  \BibitemOpen
  \bibfield  {author} {\bibinfo {author} {\bibfnamefont {M.}~\bibnamefont
  {Atzori~Corona}}, \bibinfo {author} {\bibfnamefont {M.}~\bibnamefont
  {Cadeddu}}, \bibinfo {author} {\bibfnamefont {N.}~\bibnamefont {Cargioli}},
  \bibinfo {author} {\bibfnamefont {F.}~\bibnamefont {Dordei}}, \bibinfo
  {author} {\bibfnamefont {C.}~\bibnamefont {Giunti}}, \bibinfo {author}
  {\bibfnamefont {Y.~F.}\ \bibnamefont {Li}}, \bibinfo {author} {\bibfnamefont
  {E.}~\bibnamefont {Picciau}}, \bibinfo {author} {\bibfnamefont {C.~A.}\
  \bibnamefont {Ternes}}, \ and\ \bibinfo {author} {\bibfnamefont {Y.~Y.}\
  \bibnamefont {Zhang}},\ }\href {\doibase 10.1007/JHEP05(2022)109} {\bibfield
  {journal} {\bibinfo  {journal} {JHEP}\ }\textbf {\bibinfo {volume} {05}},\
  \bibinfo {pages} {109} (\bibinfo {year} {2022})},\ \Eprint
  {http://arxiv.org/abs/2202.11002} {arXiv:2202.11002 [hep-ph]} \BibitemShut
  {NoStop}%
\bibitem [{\citenamefont {Colaresi}\ \emph {et~al.}(2021)\citenamefont
  {Colaresi}, \citenamefont {Collar}, \citenamefont {Hossbach}, \citenamefont
  {Kavner}, \citenamefont {Lewis}, \citenamefont {Robinson},\ and\
  \citenamefont {Yocum}}]{Colaresi:2021kus}%
  \BibitemOpen
  \bibfield  {author} {\bibinfo {author} {\bibfnamefont {J.}~\bibnamefont
  {Colaresi}}, \bibinfo {author} {\bibfnamefont {J.~I.}\ \bibnamefont
  {Collar}}, \bibinfo {author} {\bibfnamefont {T.~W.}\ \bibnamefont
  {Hossbach}}, \bibinfo {author} {\bibfnamefont {A.~R.~L.}\ \bibnamefont
  {Kavner}}, \bibinfo {author} {\bibfnamefont {C.~M.}\ \bibnamefont {Lewis}},
  \bibinfo {author} {\bibfnamefont {A.~E.}\ \bibnamefont {Robinson}}, \ and\
  \bibinfo {author} {\bibfnamefont {K.~M.}\ \bibnamefont {Yocum}},\ }\href
  {\doibase 10.1103/PhysRevD.104.072003} {\bibfield  {journal} {\bibinfo
  {journal} {Phys. Rev. D}\ }\textbf {\bibinfo {volume} {104}},\ \bibinfo
  {pages} {072003} (\bibinfo {year} {2021})},\ \Eprint
  {http://arxiv.org/abs/2108.02880} {arXiv:2108.02880 [hep-ex]} \BibitemShut
  {NoStop}%
\bibitem [{\citenamefont {Colaresi}\ \emph {et~al.}(2022)\citenamefont
  {Colaresi}, \citenamefont {Collar}, \citenamefont {Hossbach}, \citenamefont
  {Lewis},\ and\ \citenamefont {Yocum}}]{Colaresi:2022obx}%
  \BibitemOpen
  \bibfield  {author} {\bibinfo {author} {\bibfnamefont {J.}~\bibnamefont
  {Colaresi}}, \bibinfo {author} {\bibfnamefont {J.~I.}\ \bibnamefont
  {Collar}}, \bibinfo {author} {\bibfnamefont {T.~W.}\ \bibnamefont
  {Hossbach}}, \bibinfo {author} {\bibfnamefont {C.~M.}\ \bibnamefont {Lewis}},
  \ and\ \bibinfo {author} {\bibfnamefont {K.~M.}\ \bibnamefont {Yocum}},\
  }\href {\doibase 10.1103/PhysRevLett.129.211802} {\bibfield  {journal}
  {\bibinfo  {journal} {Phys. Rev. Lett.}\ }\textbf {\bibinfo {volume} {129}},\
  \bibinfo {pages} {211802} (\bibinfo {year} {2022})},\ \Eprint
  {http://arxiv.org/abs/2202.09672} {arXiv:2202.09672 [hep-ex]} \BibitemShut
  {NoStop}%
\bibitem [{\citenamefont {Bonet}\ \emph {et~al.}(2021)\citenamefont {Bonet}
  \emph {et~al.}}]{CONUS:2020skt}%
  \BibitemOpen
  \bibfield  {author} {\bibinfo {author} {\bibfnamefont {H.}~\bibnamefont
  {Bonet}} \emph {et~al.} (\bibinfo {collaboration} {CONUS}),\ }\href {\doibase
  10.1103/PhysRevLett.126.041804} {\bibfield  {journal} {\bibinfo  {journal}
  {Phys. Rev. Lett.}\ }\textbf {\bibinfo {volume} {126}},\ \bibinfo {pages}
  {041804} (\bibinfo {year} {2021})},\ \Eprint
  {http://arxiv.org/abs/2011.00210} {arXiv:2011.00210 [hep-ex]} \BibitemShut
  {NoStop}%
\bibitem [{\citenamefont {Bonet}\ \emph {et~al.}(2022)\citenamefont {Bonet}
  \emph {et~al.}}]{CONUS:2021dwh}%
  \BibitemOpen
  \bibfield  {author} {\bibinfo {author} {\bibfnamefont {H.}~\bibnamefont
  {Bonet}} \emph {et~al.} (\bibinfo {collaboration} {CONUS}),\ }\href {\doibase
  10.1007/JHEP05(2022)085} {\bibfield  {journal} {\bibinfo  {journal} {JHEP}\
  }\textbf {\bibinfo {volume} {05}},\ \bibinfo {pages} {085} (\bibinfo {year}
  {2022})},\ \Eprint {http://arxiv.org/abs/2110.02174} {arXiv:2110.02174
  [hep-ph]} \BibitemShut {NoStop}%
\bibitem [{\citenamefont {Maneschg}(2023)}]{conus_m7}%
  \BibitemOpen
  \bibfield  {author} {\bibinfo {author} {\bibfnamefont {W.}~\bibnamefont
  {Maneschg}},\ }\href@noop {} {\enquote {\bibinfo {title} {{Recent results
  from the CONUS experiment}},}\ }\bibinfo {howpublished}
  {\url{https://indico.cern.ch/event/1215362/contributions/5300024/}} (\bibinfo
  {year} {2023})\BibitemShut {NoStop}%
\bibitem [{\citenamefont {Verhaaren}(2022)}]{VerhaarenSlides}%
  \BibitemOpen
  \bibfield  {author} {\bibinfo {author} {\bibfnamefont {C.}~\bibnamefont
  {Verhaaren}},\ }\href {https://sites.google.com/view/hepastroresultsforum/}
  {\enquote {\bibinfo {title} {{Overview of X17 Results from ATOMKI}},}\ }
  (\bibinfo {year} {2022})\BibitemShut {NoStop}%
\bibitem [{\citenamefont {Firak}\ \emph {et~al.}(2020)\citenamefont {Firak}
  \emph {et~al.}}]{Firak:2020eil}%
  \BibitemOpen
  \bibfield  {author} {\bibinfo {author} {\bibfnamefont {D.~S.}\ \bibnamefont
  {Firak}} \emph {et~al.},\ }\href {\doibase 10.1051/epjconf/202023204005}
  {\bibfield  {journal} {\bibinfo  {journal} {EPJ Web Conf.}\ }\textbf
  {\bibinfo {volume} {232}},\ \bibinfo {pages} {04005} (\bibinfo {year}
  {2020})}\BibitemShut {NoStop}%
\bibitem [{\citenamefont {Walcher}(1970)}]{Walcher:1970vkv}%
  \BibitemOpen
  \bibfield  {author} {\bibinfo {author} {\bibfnamefont {T.}~\bibnamefont
  {Walcher}},\ }\href {\doibase 10.1016/0370-2693(70)90148-6} {\bibfield
  {journal} {\bibinfo  {journal} {Phys. Lett. B}\ }\textbf {\bibinfo {volume}
  {31}},\ \bibinfo {pages} {442} (\bibinfo {year} {1970})}\BibitemShut
  {NoStop}%
\bibitem [{\citenamefont {Raggi}(2016)}]{Raggi:2015noa}%
  \BibitemOpen
  \bibfield  {author} {\bibinfo {author} {\bibfnamefont {M.}~\bibnamefont
  {Raggi}} (\bibinfo {collaboration} {NA48/2}),\ }\href {\doibase
  10.1393/ncc/i2015-15132-0} {\bibfield  {journal} {\bibinfo  {journal} {Nuovo
  Cim. C}\ }\textbf {\bibinfo {volume} {38}},\ \bibinfo {pages} {132} (\bibinfo
  {year} {2016})},\ \Eprint {http://arxiv.org/abs/1508.01307} {arXiv:1508.01307
  [hep-ex]} \BibitemShut {NoStop}%
\bibitem [{\citenamefont {Leveille}(1978)}]{Leveille:1977rc}%
  \BibitemOpen
  \bibfield  {author} {\bibinfo {author} {\bibfnamefont {J.~P.}\ \bibnamefont
  {Leveille}},\ }\href {\doibase 10.1016/0550-3213(78)90051-2} {\bibfield
  {journal} {\bibinfo  {journal} {Nucl. Phys. B}\ }\textbf {\bibinfo {volume}
  {137}},\ \bibinfo {pages} {63} (\bibinfo {year} {1978})}\BibitemShut
  {NoStop}%
\bibitem [{\citenamefont {Fan}\ \emph {et~al.}(2023)\citenamefont {Fan},
  \citenamefont {Myers}, \citenamefont {Sukra},\ and\ \citenamefont
  {Gabrielse}}]{Fan:2022eto}%
  \BibitemOpen
  \bibfield  {author} {\bibinfo {author} {\bibfnamefont {X.}~\bibnamefont
  {Fan}}, \bibinfo {author} {\bibfnamefont {T.~G.}\ \bibnamefont {Myers}},
  \bibinfo {author} {\bibfnamefont {B.~A.~D.}\ \bibnamefont {Sukra}}, \ and\
  \bibinfo {author} {\bibfnamefont {G.}~\bibnamefont {Gabrielse}},\ }\href
  {\doibase 10.1103/PhysRevLett.130.071801} {\bibfield  {journal} {\bibinfo
  {journal} {Phys. Rev. Lett.}\ }\textbf {\bibinfo {volume} {130}},\ \bibinfo
  {pages} {071801} (\bibinfo {year} {2023})},\ \Eprint
  {http://arxiv.org/abs/2209.13084} {arXiv:2209.13084 [physics.atom-ph]}
  \BibitemShut {NoStop}%
\bibitem [{\citenamefont {Morel}\ \emph {et~al.}(2020)\citenamefont {Morel},
  \citenamefont {Yao}, \citenamefont {Clad\'e},\ and\ \citenamefont
  {Guellati-Kh\'elifa}}]{Morel:2020dww}%
  \BibitemOpen
  \bibfield  {author} {\bibinfo {author} {\bibfnamefont {L.}~\bibnamefont
  {Morel}}, \bibinfo {author} {\bibfnamefont {Z.}~\bibnamefont {Yao}}, \bibinfo
  {author} {\bibfnamefont {P.}~\bibnamefont {Clad\'e}}, \ and\ \bibinfo
  {author} {\bibfnamefont {S.}~\bibnamefont {Guellati-Kh\'elifa}},\ }\href
  {\doibase 10.1038/s41586-020-2964-7} {\bibfield  {journal} {\bibinfo
  {journal} {Nature}\ }\textbf {\bibinfo {volume} {588}},\ \bibinfo {pages}
  {61} (\bibinfo {year} {2020})}\BibitemShut {NoStop}%
\bibitem [{\citenamefont {Parker}\ \emph {et~al.}(2018)\citenamefont {Parker},
  \citenamefont {Yu}, \citenamefont {Zhong}, \citenamefont {Estey},\ and\
  \citenamefont {M\"uller}}]{Parker:2018vye}%
  \BibitemOpen
  \bibfield  {author} {\bibinfo {author} {\bibfnamefont {R.~H.}\ \bibnamefont
  {Parker}}, \bibinfo {author} {\bibfnamefont {C.}~\bibnamefont {Yu}}, \bibinfo
  {author} {\bibfnamefont {W.}~\bibnamefont {Zhong}}, \bibinfo {author}
  {\bibfnamefont {B.}~\bibnamefont {Estey}}, \ and\ \bibinfo {author}
  {\bibfnamefont {H.}~\bibnamefont {M\"uller}},\ }\href {\doibase
  10.1126/science.aap7706} {\bibfield  {journal} {\bibinfo  {journal}
  {Science}\ }\textbf {\bibinfo {volume} {360}},\ \bibinfo {pages} {191}
  (\bibinfo {year} {2018})},\ \Eprint {http://arxiv.org/abs/1812.04130}
  {arXiv:1812.04130 [physics.atom-ph]} \BibitemShut {NoStop}%
\bibitem [{\citenamefont {Banerjee}\ \emph {et~al.}(2020)\citenamefont
  {Banerjee} \emph {et~al.}}]{NA64:2019auh}%
  \BibitemOpen
  \bibfield  {author} {\bibinfo {author} {\bibfnamefont {D.}~\bibnamefont
  {Banerjee}} \emph {et~al.} (\bibinfo {collaboration} {NA64}),\ }\href
  {\doibase 10.1103/PhysRevD.101.071101} {\bibfield  {journal} {\bibinfo
  {journal} {Phys. Rev. D}\ }\textbf {\bibinfo {volume} {101}},\ \bibinfo
  {pages} {071101} (\bibinfo {year} {2020})},\ \Eprint
  {http://arxiv.org/abs/1912.11389} {arXiv:1912.11389 [hep-ex]} \BibitemShut
  {NoStop}%
\bibitem [{\citenamefont {Bjorken}\ \emph {et~al.}(1988)\citenamefont
  {Bjorken}, \citenamefont {Ecklund}, \citenamefont {Nelson}, \citenamefont
  {Abashian}, \citenamefont {Church}, \citenamefont {Lu}, \citenamefont {Mo},
  \citenamefont {Nunamaker},\ and\ \citenamefont {Rassmann}}]{Bjorken:1988as}%
  \BibitemOpen
  \bibfield  {author} {\bibinfo {author} {\bibfnamefont {J.~D.}\ \bibnamefont
  {Bjorken}}, \bibinfo {author} {\bibfnamefont {S.}~\bibnamefont {Ecklund}},
  \bibinfo {author} {\bibfnamefont {W.~R.}\ \bibnamefont {Nelson}}, \bibinfo
  {author} {\bibfnamefont {A.}~\bibnamefont {Abashian}}, \bibinfo {author}
  {\bibfnamefont {C.}~\bibnamefont {Church}}, \bibinfo {author} {\bibfnamefont
  {B.}~\bibnamefont {Lu}}, \bibinfo {author} {\bibfnamefont {L.~W.}\
  \bibnamefont {Mo}}, \bibinfo {author} {\bibfnamefont {T.~A.}\ \bibnamefont
  {Nunamaker}}, \ and\ \bibinfo {author} {\bibfnamefont {P.}~\bibnamefont
  {Rassmann}},\ }\href {\doibase 10.1103/PhysRevD.38.3375} {\bibfield
  {journal} {\bibinfo  {journal} {Phys. Rev. D}\ }\textbf {\bibinfo {volume}
  {38}},\ \bibinfo {pages} {3375} (\bibinfo {year} {1988})}\BibitemShut
  {NoStop}%
\bibitem [{\citenamefont {Bjorken}\ \emph {et~al.}(2009)\citenamefont
  {Bjorken}, \citenamefont {Essig}, \citenamefont {Schuster},\ and\
  \citenamefont {Toro}}]{Bjorken:2009mm}%
  \BibitemOpen
  \bibfield  {author} {\bibinfo {author} {\bibfnamefont {J.~D.}\ \bibnamefont
  {Bjorken}}, \bibinfo {author} {\bibfnamefont {R.}~\bibnamefont {Essig}},
  \bibinfo {author} {\bibfnamefont {P.}~\bibnamefont {Schuster}}, \ and\
  \bibinfo {author} {\bibfnamefont {N.}~\bibnamefont {Toro}},\ }\href {\doibase
  10.1103/PhysRevD.80.075018} {\bibfield  {journal} {\bibinfo  {journal} {Phys.
  Rev. D}\ }\textbf {\bibinfo {volume} {80}},\ \bibinfo {pages} {075018}
  (\bibinfo {year} {2009})},\ \Eprint {http://arxiv.org/abs/0906.0580}
  {arXiv:0906.0580 [hep-ph]} \BibitemShut {NoStop}%
\bibitem [{\citenamefont {Dent}\ \emph {et~al.}(2012)\citenamefont {Dent},
  \citenamefont {Ferrer},\ and\ \citenamefont {Krauss}}]{Dent:2012mx}%
  \BibitemOpen
  \bibfield  {author} {\bibinfo {author} {\bibfnamefont {J.~B.}\ \bibnamefont
  {Dent}}, \bibinfo {author} {\bibfnamefont {F.}~\bibnamefont {Ferrer}}, \ and\
  \bibinfo {author} {\bibfnamefont {L.~M.}\ \bibnamefont {Krauss}},\
  }\href@noop {} {\  (\bibinfo {year} {2012})},\ \Eprint
  {http://arxiv.org/abs/1201.2683} {arXiv:1201.2683 [astro-ph.CO]} \BibitemShut
  {NoStop}%
\bibitem [{\citenamefont {Dreiner}\ \emph {et~al.}(2014)\citenamefont
  {Dreiner}, \citenamefont {Fortin}, \citenamefont {Hanhart},\ and\
  \citenamefont {Ubaldi}}]{Dreiner:2013mua}%
  \BibitemOpen
  \bibfield  {author} {\bibinfo {author} {\bibfnamefont {H.~K.}\ \bibnamefont
  {Dreiner}}, \bibinfo {author} {\bibfnamefont {J.-F.}\ \bibnamefont {Fortin}},
  \bibinfo {author} {\bibfnamefont {C.}~\bibnamefont {Hanhart}}, \ and\
  \bibinfo {author} {\bibfnamefont {L.}~\bibnamefont {Ubaldi}},\ }\href
  {\doibase 10.1103/PhysRevD.89.105015} {\bibfield  {journal} {\bibinfo
  {journal} {Phys. Rev. D}\ }\textbf {\bibinfo {volume} {89}},\ \bibinfo
  {pages} {105015} (\bibinfo {year} {2014})},\ \Eprint
  {http://arxiv.org/abs/1310.3826} {arXiv:1310.3826 [hep-ph]} \BibitemShut
  {NoStop}%
\bibitem [{\citenamefont {Chang}\ \emph {et~al.}(2018)\citenamefont {Chang},
  \citenamefont {Essig},\ and\ \citenamefont {McDermott}}]{Chang:2018rso}%
  \BibitemOpen
  \bibfield  {author} {\bibinfo {author} {\bibfnamefont {J.~H.}\ \bibnamefont
  {Chang}}, \bibinfo {author} {\bibfnamefont {R.}~\bibnamefont {Essig}}, \ and\
  \bibinfo {author} {\bibfnamefont {S.~D.}\ \bibnamefont {McDermott}},\ }\href
  {\doibase 10.1007/JHEP09(2018)051} {\bibfield  {journal} {\bibinfo  {journal}
  {JHEP}\ }\textbf {\bibinfo {volume} {09}},\ \bibinfo {pages} {051} (\bibinfo
  {year} {2018})},\ \Eprint {http://arxiv.org/abs/1803.00993} {arXiv:1803.00993
  [hep-ph]} \BibitemShut {NoStop}%
\bibitem [{\citenamefont {Kazanas}\ \emph {et~al.}(2014)\citenamefont
  {Kazanas}, \citenamefont {Mohapatra}, \citenamefont {Nussinov}, \citenamefont
  {Teplitz},\ and\ \citenamefont {Zhang}}]{Kazanas:2014mca}%
  \BibitemOpen
  \bibfield  {author} {\bibinfo {author} {\bibfnamefont {D.}~\bibnamefont
  {Kazanas}}, \bibinfo {author} {\bibfnamefont {R.~N.}\ \bibnamefont
  {Mohapatra}}, \bibinfo {author} {\bibfnamefont {S.}~\bibnamefont {Nussinov}},
  \bibinfo {author} {\bibfnamefont {V.~L.}\ \bibnamefont {Teplitz}}, \ and\
  \bibinfo {author} {\bibfnamefont {Y.}~\bibnamefont {Zhang}},\ }\href
  {\doibase 10.1016/j.nuclphysb.2014.11.009} {\bibfield  {journal} {\bibinfo
  {journal} {Nucl. Phys. B}\ }\textbf {\bibinfo {volume} {890}},\ \bibinfo
  {pages} {17} (\bibinfo {year} {2014})},\ \Eprint
  {http://arxiv.org/abs/1410.0221} {arXiv:1410.0221 [hep-ph]} \BibitemShut
  {NoStop}%
\bibitem [{\citenamefont {Coloma}\ \emph {et~al.}(2022)\citenamefont {Coloma},
  \citenamefont {Esteban}, \citenamefont {Gonzalez-Garcia}, \citenamefont
  {Larizgoitia}, \citenamefont {Monrabal},\ and\ \citenamefont
  {Palomares-Ruiz}}]{Coloma:2022avw}%
  \BibitemOpen
  \bibfield  {author} {\bibinfo {author} {\bibfnamefont {P.}~\bibnamefont
  {Coloma}}, \bibinfo {author} {\bibfnamefont {I.}~\bibnamefont {Esteban}},
  \bibinfo {author} {\bibfnamefont {M.~C.}\ \bibnamefont {Gonzalez-Garcia}},
  \bibinfo {author} {\bibfnamefont {L.}~\bibnamefont {Larizgoitia}}, \bibinfo
  {author} {\bibfnamefont {F.}~\bibnamefont {Monrabal}}, \ and\ \bibinfo
  {author} {\bibfnamefont {S.}~\bibnamefont {Palomares-Ruiz}},\ }\href
  {\doibase 10.1007/JHEP05(2022)037} {\bibfield  {journal} {\bibinfo  {journal}
  {JHEP}\ }\textbf {\bibinfo {volume} {05}},\ \bibinfo {pages} {037} (\bibinfo
  {year} {2022})},\ \Eprint {http://arxiv.org/abs/2202.10829} {arXiv:2202.10829
  [hep-ph]} \BibitemShut {NoStop}%
\bibitem [{\citenamefont {Kling}(2020)}]{Kling:2020iar}%
  \BibitemOpen
  \bibfield  {author} {\bibinfo {author} {\bibfnamefont {F.}~\bibnamefont
  {Kling}},\ }\href {\doibase 10.1103/PhysRevD.102.015007} {\bibfield
  {journal} {\bibinfo  {journal} {Phys. Rev. D}\ }\textbf {\bibinfo {volume}
  {102}},\ \bibinfo {pages} {015007} (\bibinfo {year} {2020})},\ \Eprint
  {http://arxiv.org/abs/2005.03594} {arXiv:2005.03594 [hep-ph]} \BibitemShut
  {NoStop}%
\bibitem [{\citenamefont {Aguilar-Arevalo}\ \emph {et~al.}(2020)\citenamefont
  {Aguilar-Arevalo} \emph {et~al.}}]{CONNIE:2019xid}%
  \BibitemOpen
  \bibfield  {author} {\bibinfo {author} {\bibfnamefont {A.}~\bibnamefont
  {Aguilar-Arevalo}} \emph {et~al.} (\bibinfo {collaboration} {CONNIE}),\
  }\href {\doibase 10.1007/JHEP04(2020)054} {\bibfield  {journal} {\bibinfo
  {journal} {JHEP}\ }\textbf {\bibinfo {volume} {04}},\ \bibinfo {pages} {054}
  (\bibinfo {year} {2020})},\ \Eprint {http://arxiv.org/abs/1910.04951}
  {arXiv:1910.04951 [hep-ex]} \BibitemShut {NoStop}%
\bibitem [{\citenamefont {Castro}\ and\ \citenamefont
  {Quintero}(2021)}]{Castro:2021gdf}%
  \BibitemOpen
  \bibfield  {author} {\bibinfo {author} {\bibfnamefont {G.~L.}\ \bibnamefont
  {Castro}}\ and\ \bibinfo {author} {\bibfnamefont {N.}~\bibnamefont
  {Quintero}},\ }\href {\doibase 10.1103/PhysRevD.103.093002} {\bibfield
  {journal} {\bibinfo  {journal} {Phys. Rev. D}\ }\textbf {\bibinfo {volume}
  {103}},\ \bibinfo {pages} {093002} (\bibinfo {year} {2021})},\ \Eprint
  {http://arxiv.org/abs/2101.01865} {arXiv:2101.01865 [hep-ph]} \BibitemShut
  {NoStop}%
\bibitem [{\citenamefont {Wolfenstein}(1978)}]{Wolfenstein:1977ue}%
  \BibitemOpen
  \bibfield  {author} {\bibinfo {author} {\bibfnamefont {L.}~\bibnamefont
  {Wolfenstein}},\ }\href {\doibase 10.1103/PhysRevD.17.2369} {\bibfield
  {journal} {\bibinfo  {journal} {Phys. Rev. D}\ }\textbf {\bibinfo {volume}
  {17}},\ \bibinfo {pages} {2369} (\bibinfo {year} {1978})}\BibitemShut
  {NoStop}%
\bibitem [{Pro(2019)}]{Proceedings:2019qno}%
  \BibitemOpen
  \href {\doibase 10.21468/SciPostPhysProc.2.001} {\emph {\bibinfo {title}
  {{Neutrino Non-Standard Interactions: A Status Report}}}},\ Vol.~\bibinfo
  {volume} {2}\ (\bibinfo {year} {2019})\ \Eprint
  {http://arxiv.org/abs/1907.00991} {arXiv:1907.00991 [hep-ph]} \BibitemShut
  {NoStop}%
\bibitem [{\citenamefont {Esteban}\ \emph {et~al.}(2018)\citenamefont
  {Esteban}, \citenamefont {Gonzalez-Garcia}, \citenamefont {Maltoni},
  \citenamefont {Martinez-Soler},\ and\ \citenamefont
  {Salvado}}]{Esteban:2018ppq}%
  \BibitemOpen
  \bibfield  {author} {\bibinfo {author} {\bibfnamefont {I.}~\bibnamefont
  {Esteban}}, \bibinfo {author} {\bibfnamefont {M.~C.}\ \bibnamefont
  {Gonzalez-Garcia}}, \bibinfo {author} {\bibfnamefont {M.}~\bibnamefont
  {Maltoni}}, \bibinfo {author} {\bibfnamefont {I.}~\bibnamefont
  {Martinez-Soler}}, \ and\ \bibinfo {author} {\bibfnamefont {J.}~\bibnamefont
  {Salvado}},\ }\href {\doibase 10.1007/JHEP08(2018)180} {\bibfield  {journal}
  {\bibinfo  {journal} {JHEP}\ }\textbf {\bibinfo {volume} {08}},\ \bibinfo
  {pages} {180} (\bibinfo {year} {2018})},\ \bibinfo {note} {[Addendum: JHEP
  12, 152 (2020)]},\ \Eprint {http://arxiv.org/abs/1805.04530}
  {arXiv:1805.04530 [hep-ph]} \BibitemShut {NoStop}%
\bibitem [{\citenamefont {Abusleme}\ \emph {et~al.}(2022)\citenamefont
  {Abusleme} \emph {et~al.}}]{JUNO:2022mxj}%
  \BibitemOpen
  \bibfield  {author} {\bibinfo {author} {\bibfnamefont {A.}~\bibnamefont
  {Abusleme}} \emph {et~al.} (\bibinfo {collaboration} {JUNO}),\ }\href
  {\doibase 10.1088/1674-1137/ac8bc9} {\bibfield  {journal} {\bibinfo
  {journal} {Chin. Phys. C}\ }\textbf {\bibinfo {volume} {46}},\ \bibinfo
  {pages} {123001} (\bibinfo {year} {2022})},\ \Eprint
  {http://arxiv.org/abs/2204.13249} {arXiv:2204.13249 [hep-ex]} \BibitemShut
  {NoStop}%
\bibitem [{\citenamefont {Capozzi}\ \emph {et~al.}(2019)\citenamefont
  {Capozzi}, \citenamefont {Li}, \citenamefont {Zhu},\ and\ \citenamefont
  {Beacom}}]{Capozzi:2018dat}%
  \BibitemOpen
  \bibfield  {author} {\bibinfo {author} {\bibfnamefont {F.}~\bibnamefont
  {Capozzi}}, \bibinfo {author} {\bibfnamefont {S.~W.}\ \bibnamefont {Li}},
  \bibinfo {author} {\bibfnamefont {G.}~\bibnamefont {Zhu}}, \ and\ \bibinfo
  {author} {\bibfnamefont {J.~F.}\ \bibnamefont {Beacom}},\ }\href {\doibase
  10.1103/PhysRevLett.123.131803} {\bibfield  {journal} {\bibinfo  {journal}
  {Phys. Rev. Lett.}\ }\textbf {\bibinfo {volume} {123}},\ \bibinfo {pages}
  {131803} (\bibinfo {year} {2019})},\ \Eprint
  {http://arxiv.org/abs/1808.08232} {arXiv:1808.08232 [hep-ph]} \BibitemShut
  {NoStop}%
\bibitem [{\citenamefont {Abi}\ \emph {et~al.}(2020)\citenamefont {Abi} \emph
  {et~al.}}]{DUNE:2020ypp}%
  \BibitemOpen
  \bibfield  {author} {\bibinfo {author} {\bibfnamefont {B.}~\bibnamefont
  {Abi}} \emph {et~al.} (\bibinfo {collaboration} {DUNE}),\ }\href@noop {} {\
  (\bibinfo {year} {2020})},\ \Eprint {http://arxiv.org/abs/2002.03005}
  {arXiv:2002.03005 [hep-ex]} \BibitemShut {NoStop}%
\bibitem [{\citenamefont {Abdullah}\ \emph {et~al.}(2022)\citenamefont
  {Abdullah} \emph {et~al.}}]{Abdullah:2022zue}%
  \BibitemOpen
  \bibfield  {author} {\bibinfo {author} {\bibfnamefont {M.}~\bibnamefont
  {Abdullah}} \emph {et~al.},\ }\href@noop {} {\  (\bibinfo {year} {2022})},\
  \Eprint {http://arxiv.org/abs/2203.07361} {arXiv:2203.07361 [hep-ph]}
  \BibitemShut {NoStop}%
\bibitem [{\citenamefont {Fernandez-Moroni}\ \emph {et~al.}(2022)\citenamefont
  {Fernandez-Moroni}, \citenamefont {Harnik}, \citenamefont {Machado},
  \citenamefont {Martinez-Soler}, \citenamefont {Perez-Gonzalez}, \citenamefont
  {Rodrigues},\ and\ \citenamefont
  {Rosauro-Alcaraz}}]{Fernandez-Moroni:2021nap}%
  \BibitemOpen
  \bibfield  {author} {\bibinfo {author} {\bibfnamefont {G.}~\bibnamefont
  {Fernandez-Moroni}}, \bibinfo {author} {\bibfnamefont {R.}~\bibnamefont
  {Harnik}}, \bibinfo {author} {\bibfnamefont {P.~A.~N.}\ \bibnamefont
  {Machado}}, \bibinfo {author} {\bibfnamefont {I.}~\bibnamefont
  {Martinez-Soler}}, \bibinfo {author} {\bibfnamefont {Y.~F.}\ \bibnamefont
  {Perez-Gonzalez}}, \bibinfo {author} {\bibfnamefont {D.}~\bibnamefont
  {Rodrigues}}, \ and\ \bibinfo {author} {\bibfnamefont {S.}~\bibnamefont
  {Rosauro-Alcaraz}},\ }\href {\doibase 10.1007/JHEP02(2022)127} {\bibfield
  {journal} {\bibinfo  {journal} {JHEP}\ }\textbf {\bibinfo {volume} {02}},\
  \bibinfo {pages} {127} (\bibinfo {year} {2022})},\ \Eprint
  {http://arxiv.org/abs/2108.07310} {arXiv:2108.07310 [hep-ph]} \BibitemShut
  {NoStop}%
\bibitem [{\citenamefont {Ilten}\ \emph {et~al.}(2015)\citenamefont {Ilten},
  \citenamefont {Thaler}, \citenamefont {Williams},\ and\ \citenamefont
  {Xue}}]{Ilten:2015hya}%
  \BibitemOpen
  \bibfield  {author} {\bibinfo {author} {\bibfnamefont {P.}~\bibnamefont
  {Ilten}}, \bibinfo {author} {\bibfnamefont {J.}~\bibnamefont {Thaler}},
  \bibinfo {author} {\bibfnamefont {M.}~\bibnamefont {Williams}}, \ and\
  \bibinfo {author} {\bibfnamefont {W.}~\bibnamefont {Xue}},\ }\href {\doibase
  10.1103/PhysRevD.92.115017} {\bibfield  {journal} {\bibinfo  {journal} {Phys.
  Rev. D}\ }\textbf {\bibinfo {volume} {92}},\ \bibinfo {pages} {115017}
  (\bibinfo {year} {2015})},\ \Eprint {http://arxiv.org/abs/1509.06765}
  {arXiv:1509.06765 [hep-ph]} \BibitemShut {NoStop}%
\bibitem [{\citenamefont {Apyan}\ \emph {et~al.}(2022)\citenamefont {Apyan}
  \emph {et~al.}}]{Apyan:2022tsd}%
  \BibitemOpen
  \bibfield  {author} {\bibinfo {author} {\bibfnamefont {A.}~\bibnamefont
  {Apyan}} \emph {et~al.},\ }in\ \href@noop {} {\emph {\bibinfo {booktitle}
  {{Snowmass 2021}}}}\ (\bibinfo {year} {2022})\ \Eprint
  {http://arxiv.org/abs/2203.08322} {arXiv:2203.08322 [hep-ex]} \BibitemShut
  {NoStop}%
\bibitem [{\citenamefont {Ariga}\ \emph {et~al.}(2019)\citenamefont {Ariga}
  \emph {et~al.}}]{FASER:2018eoc}%
  \BibitemOpen
  \bibfield  {author} {\bibinfo {author} {\bibfnamefont {A.}~\bibnamefont
  {Ariga}} \emph {et~al.} (\bibinfo {collaboration} {FASER}),\ }\href {\doibase
  10.1103/PhysRevD.99.095011} {\bibfield  {journal} {\bibinfo  {journal} {Phys.
  Rev. D}\ }\textbf {\bibinfo {volume} {99}},\ \bibinfo {pages} {095011}
  (\bibinfo {year} {2019})},\ \Eprint {http://arxiv.org/abs/1811.12522}
  {arXiv:1811.12522 [hep-ph]} \BibitemShut {NoStop}%
\bibitem [{\citenamefont {Echenard}\ \emph {et~al.}(2015)\citenamefont
  {Echenard}, \citenamefont {Essig},\ and\ \citenamefont
  {Zhong}}]{Echenard:2014lma}%
  \BibitemOpen
  \bibfield  {author} {\bibinfo {author} {\bibfnamefont {B.}~\bibnamefont
  {Echenard}}, \bibinfo {author} {\bibfnamefont {R.}~\bibnamefont {Essig}}, \
  and\ \bibinfo {author} {\bibfnamefont {Y.-M.}\ \bibnamefont {Zhong}},\ }\href
  {\doibase 10.1007/JHEP01(2015)113} {\bibfield  {journal} {\bibinfo  {journal}
  {JHEP}\ }\textbf {\bibinfo {volume} {01}},\ \bibinfo {pages} {113} (\bibinfo
  {year} {2015})},\ \Eprint {http://arxiv.org/abs/1411.1770} {arXiv:1411.1770
  [hep-ph]} \BibitemShut {NoStop}%
\bibitem [{\citenamefont {Fileviez~Perez}\ and\ \citenamefont
  {Wise}(2010)}]{FileviezPerez:2010gw}%
  \BibitemOpen
  \bibfield  {author} {\bibinfo {author} {\bibfnamefont {P.}~\bibnamefont
  {Fileviez~Perez}}\ and\ \bibinfo {author} {\bibfnamefont {M.~B.}\
  \bibnamefont {Wise}},\ }\href {\doibase 10.1103/PhysRevD.82.079901}
  {\bibfield  {journal} {\bibinfo  {journal} {Phys. Rev. D}\ }\textbf {\bibinfo
  {volume} {82}},\ \bibinfo {pages} {011901} (\bibinfo {year} {2010})},\
  \bibinfo {note} {[Erratum: Phys.Rev.D 82, 079901 (2010)]},\ \Eprint
  {http://arxiv.org/abs/1002.1754} {arXiv:1002.1754 [hep-ph]} \BibitemShut
  {NoStop}%
\bibitem [{\citenamefont {Dulaney}\ \emph {et~al.}(2011)\citenamefont
  {Dulaney}, \citenamefont {Fileviez~Perez},\ and\ \citenamefont
  {Wise}}]{Dulaney:2010dj}%
  \BibitemOpen
  \bibfield  {author} {\bibinfo {author} {\bibfnamefont {T.~R.}\ \bibnamefont
  {Dulaney}}, \bibinfo {author} {\bibfnamefont {P.}~\bibnamefont
  {Fileviez~Perez}}, \ and\ \bibinfo {author} {\bibfnamefont {M.~B.}\
  \bibnamefont {Wise}},\ }\href {\doibase 10.1103/PhysRevD.83.023520}
  {\bibfield  {journal} {\bibinfo  {journal} {Phys. Rev. D}\ }\textbf {\bibinfo
  {volume} {83}},\ \bibinfo {pages} {023520} (\bibinfo {year} {2011})},\
  \Eprint {http://arxiv.org/abs/1005.0617} {arXiv:1005.0617 [hep-ph]}
  \BibitemShut {NoStop}%
\bibitem [{\citenamefont {Fileviez~Perez}\ and\ \citenamefont
  {Wise}(2011{\natexlab{a}})}]{FileviezPerez:2011dg}%
  \BibitemOpen
  \bibfield  {author} {\bibinfo {author} {\bibfnamefont {P.}~\bibnamefont
  {Fileviez~Perez}}\ and\ \bibinfo {author} {\bibfnamefont {M.~B.}\
  \bibnamefont {Wise}},\ }\href {\doibase 10.1103/PhysRevD.84.055015}
  {\bibfield  {journal} {\bibinfo  {journal} {Phys. Rev. D}\ }\textbf {\bibinfo
  {volume} {84}},\ \bibinfo {pages} {055015} (\bibinfo {year}
  {2011}{\natexlab{a}})},\ \Eprint {http://arxiv.org/abs/1105.3190}
  {arXiv:1105.3190 [hep-ph]} \BibitemShut {NoStop}%
\bibitem [{\citenamefont {Fileviez~Perez}\ and\ \citenamefont
  {Wise}(2011{\natexlab{b}})}]{FileviezPerez:2011pt}%
  \BibitemOpen
  \bibfield  {author} {\bibinfo {author} {\bibfnamefont {P.}~\bibnamefont
  {Fileviez~Perez}}\ and\ \bibinfo {author} {\bibfnamefont {M.~B.}\
  \bibnamefont {Wise}},\ }\href {\doibase 10.1007/JHEP08(2011)068} {\bibfield
  {journal} {\bibinfo  {journal} {JHEP}\ }\textbf {\bibinfo {volume} {08}},\
  \bibinfo {pages} {068} (\bibinfo {year} {2011}{\natexlab{b}})},\ \Eprint
  {http://arxiv.org/abs/1106.0343} {arXiv:1106.0343 [hep-ph]} \BibitemShut
  {NoStop}%
\bibitem [{\citenamefont {Duerr}\ \emph {et~al.}(2013)\citenamefont {Duerr},
  \citenamefont {Fileviez~Perez},\ and\ \citenamefont {Wise}}]{Duerr:2013dza}%
  \BibitemOpen
  \bibfield  {author} {\bibinfo {author} {\bibfnamefont {M.}~\bibnamefont
  {Duerr}}, \bibinfo {author} {\bibfnamefont {P.}~\bibnamefont
  {Fileviez~Perez}}, \ and\ \bibinfo {author} {\bibfnamefont {M.~B.}\
  \bibnamefont {Wise}},\ }\href {\doibase 10.1103/PhysRevLett.110.231801}
  {\bibfield  {journal} {\bibinfo  {journal} {Phys. Rev. Lett.}\ }\textbf
  {\bibinfo {volume} {110}},\ \bibinfo {pages} {231801} (\bibinfo {year}
  {2013})},\ \Eprint {http://arxiv.org/abs/1304.0576} {arXiv:1304.0576
  [hep-ph]} \BibitemShut {NoStop}%
\bibitem [{\citenamefont {Arnold}\ \emph {et~al.}(2013)\citenamefont {Arnold},
  \citenamefont {Fileviez~P\'erez}, \citenamefont {Fornal},\ and\ \citenamefont
  {Spinner}}]{Arnold:2013qja}%
  \BibitemOpen
  \bibfield  {author} {\bibinfo {author} {\bibfnamefont {J.~M.}\ \bibnamefont
  {Arnold}}, \bibinfo {author} {\bibfnamefont {P.}~\bibnamefont
  {Fileviez~P\'erez}}, \bibinfo {author} {\bibfnamefont {B.}~\bibnamefont
  {Fornal}}, \ and\ \bibinfo {author} {\bibfnamefont {S.}~\bibnamefont
  {Spinner}},\ }\href {\doibase 10.1103/PhysRevD.88.115009} {\bibfield
  {journal} {\bibinfo  {journal} {Phys. Rev. D}\ }\textbf {\bibinfo {volume}
  {88}},\ \bibinfo {pages} {115009} (\bibinfo {year} {2013})},\ \Eprint
  {http://arxiv.org/abs/1310.7052} {arXiv:1310.7052 [hep-ph]} \BibitemShut
  {NoStop}%
\bibitem [{\citenamefont {Fileviez~Perez}\ \emph {et~al.}(2014)\citenamefont
  {Fileviez~Perez}, \citenamefont {Ohmer},\ and\ \citenamefont
  {Patel}}]{FileviezPerez:2014lnj}%
  \BibitemOpen
  \bibfield  {author} {\bibinfo {author} {\bibfnamefont {P.}~\bibnamefont
  {Fileviez~Perez}}, \bibinfo {author} {\bibfnamefont {S.}~\bibnamefont
  {Ohmer}}, \ and\ \bibinfo {author} {\bibfnamefont {H.~H.}\ \bibnamefont
  {Patel}},\ }\href {\doibase 10.1016/j.physletb.2014.06.057} {\bibfield
  {journal} {\bibinfo  {journal} {Phys. Lett. B}\ }\textbf {\bibinfo {volume}
  {735}},\ \bibinfo {pages} {283} (\bibinfo {year} {2014})},\ \Eprint
  {http://arxiv.org/abs/1403.8029} {arXiv:1403.8029 [hep-ph]} \BibitemShut
  {NoStop}%
\bibitem [{\citenamefont {Duerr}\ and\ \citenamefont
  {Fileviez~Perez}(2015)}]{Duerr:2014wra}%
  \BibitemOpen
  \bibfield  {author} {\bibinfo {author} {\bibfnamefont {M.}~\bibnamefont
  {Duerr}}\ and\ \bibinfo {author} {\bibfnamefont {P.}~\bibnamefont
  {Fileviez~Perez}},\ }\href {\doibase 10.1103/PhysRevD.91.095001} {\bibfield
  {journal} {\bibinfo  {journal} {Phys. Rev. D}\ }\textbf {\bibinfo {volume}
  {91}},\ \bibinfo {pages} {095001} (\bibinfo {year} {2015})},\ \Eprint
  {http://arxiv.org/abs/1409.8165} {arXiv:1409.8165 [hep-ph]} \BibitemShut
  {NoStop}%
\bibitem [{\citenamefont {Farzan}\ and\ \citenamefont
  {Heeck}(2016)}]{Farzan:2016wym}%
  \BibitemOpen
  \bibfield  {author} {\bibinfo {author} {\bibfnamefont {Y.}~\bibnamefont
  {Farzan}}\ and\ \bibinfo {author} {\bibfnamefont {J.}~\bibnamefont {Heeck}},\
  }\href {\doibase 10.1103/PhysRevD.94.053010} {\bibfield  {journal} {\bibinfo
  {journal} {Phys. Rev. D}\ }\textbf {\bibinfo {volume} {94}},\ \bibinfo
  {pages} {053010} (\bibinfo {year} {2016})},\ \Eprint
  {http://arxiv.org/abs/1607.07616} {arXiv:1607.07616 [hep-ph]} \BibitemShut
  {NoStop}%
\bibitem [{\citenamefont {Workman}\ \emph {et~al.}(2022)\citenamefont {Workman}
  \emph {et~al.}}]{ParticleDataGroup:2022pth}%
  \BibitemOpen
  \bibfield  {author} {\bibinfo {author} {\bibfnamefont {R.~L.}\ \bibnamefont
  {Workman}} \emph {et~al.} (\bibinfo {collaboration} {Particle Data Group}),\
  }\href {\doibase 10.1093/ptep/ptac097} {\bibfield  {journal} {\bibinfo
  {journal} {PTEP}\ }\textbf {\bibinfo {volume} {2022}},\ \bibinfo {pages}
  {083C01} (\bibinfo {year} {2022})}\BibitemShut {NoStop}%
\bibitem [{\citenamefont {Bergstrom}\ \emph {et~al.}(2016)\citenamefont
  {Bergstrom}, \citenamefont {Gonzalez-Garcia}, \citenamefont {Maltoni},
  \citenamefont {Pena-Garay}, \citenamefont {Serenelli},\ and\ \citenamefont
  {Song}}]{Bergstrom:2016cbh}%
  \BibitemOpen
  \bibfield  {author} {\bibinfo {author} {\bibfnamefont {J.}~\bibnamefont
  {Bergstrom}}, \bibinfo {author} {\bibfnamefont {M.~C.}\ \bibnamefont
  {Gonzalez-Garcia}}, \bibinfo {author} {\bibfnamefont {M.}~\bibnamefont
  {Maltoni}}, \bibinfo {author} {\bibfnamefont {C.}~\bibnamefont {Pena-Garay}},
  \bibinfo {author} {\bibfnamefont {A.~M.}\ \bibnamefont {Serenelli}}, \ and\
  \bibinfo {author} {\bibfnamefont {N.}~\bibnamefont {Song}},\ }\href {\doibase
  10.1007/JHEP03(2016)132} {\bibfield  {journal} {\bibinfo  {journal} {JHEP}\
  }\textbf {\bibinfo {volume} {03}},\ \bibinfo {pages} {132} (\bibinfo {year}
  {2016})},\ \Eprint {http://arxiv.org/abs/1601.00972} {arXiv:1601.00972
  [hep-ph]} \BibitemShut {NoStop}%
\bibitem [{\citenamefont {Mention}\ \emph {et~al.}(2011)\citenamefont
  {Mention}, \citenamefont {Fechner}, \citenamefont {Lasserre}, \citenamefont
  {Mueller}, \citenamefont {Lhuillier}, \citenamefont {Cribier},\ and\
  \citenamefont {Letourneau}}]{Mention:2011rk}%
  \BibitemOpen
  \bibfield  {author} {\bibinfo {author} {\bibfnamefont {G.}~\bibnamefont
  {Mention}}, \bibinfo {author} {\bibfnamefont {M.}~\bibnamefont {Fechner}},
  \bibinfo {author} {\bibfnamefont {T.}~\bibnamefont {Lasserre}}, \bibinfo
  {author} {\bibfnamefont {T.~A.}\ \bibnamefont {Mueller}}, \bibinfo {author}
  {\bibfnamefont {D.}~\bibnamefont {Lhuillier}}, \bibinfo {author}
  {\bibfnamefont {M.}~\bibnamefont {Cribier}}, \ and\ \bibinfo {author}
  {\bibfnamefont {A.}~\bibnamefont {Letourneau}},\ }\href {\doibase
  10.1103/PhysRevD.83.073006} {\bibfield  {journal} {\bibinfo  {journal} {Phys.
  Rev. D}\ }\textbf {\bibinfo {volume} {83}},\ \bibinfo {pages} {073006}
  (\bibinfo {year} {2011})},\ \Eprint {http://arxiv.org/abs/1101.2755}
  {arXiv:1101.2755 [hep-ex]} \BibitemShut {NoStop}%
\bibitem [{\citenamefont {An}\ \emph {et~al.}(2017)\citenamefont {An} \emph
  {et~al.}}]{DayaBay:2017jkb}%
  \BibitemOpen
  \bibfield  {author} {\bibinfo {author} {\bibfnamefont {F.~P.}\ \bibnamefont
  {An}} \emph {et~al.} (\bibinfo {collaboration} {Daya Bay}),\ }\href {\doibase
  10.1103/PhysRevLett.118.251801} {\bibfield  {journal} {\bibinfo  {journal}
  {Phys. Rev. Lett.}\ }\textbf {\bibinfo {volume} {118}},\ \bibinfo {pages}
  {251801} (\bibinfo {year} {2017})},\ \Eprint
  {http://arxiv.org/abs/1704.01082} {arXiv:1704.01082 [hep-ex]} \BibitemShut
  {NoStop}%
\bibitem [{\citenamefont {Bak}\ \emph {et~al.}(2019)\citenamefont {Bak} \emph
  {et~al.}}]{RENO:2018pwo}%
  \BibitemOpen
  \bibfield  {author} {\bibinfo {author} {\bibfnamefont {G.}~\bibnamefont
  {Bak}} \emph {et~al.} (\bibinfo {collaboration} {RENO}),\ }\href {\doibase
  10.1103/PhysRevLett.122.232501} {\bibfield  {journal} {\bibinfo  {journal}
  {Phys. Rev. Lett.}\ }\textbf {\bibinfo {volume} {122}},\ \bibinfo {pages}
  {232501} (\bibinfo {year} {2019})},\ \Eprint
  {http://arxiv.org/abs/1806.00574} {arXiv:1806.00574 [hep-ex]} \BibitemShut
  {NoStop}%
\bibitem [{\citenamefont {Anselmann}\ \emph {et~al.}(1995)\citenamefont
  {Anselmann} \emph {et~al.}}]{GALLEX:1994rym}%
  \BibitemOpen
  \bibfield  {author} {\bibinfo {author} {\bibfnamefont {P.}~\bibnamefont
  {Anselmann}} \emph {et~al.} (\bibinfo {collaboration} {GALLEX}),\ }\href
  {\doibase 10.1016/0370-2693(94)01586-2} {\bibfield  {journal} {\bibinfo
  {journal} {Phys. Lett. B}\ }\textbf {\bibinfo {volume} {342}},\ \bibinfo
  {pages} {440} (\bibinfo {year} {1995})}\BibitemShut {NoStop}%
\bibitem [{\citenamefont {Hampel}\ \emph {et~al.}(1998)\citenamefont {Hampel}
  \emph {et~al.}}]{GALLEX:1997lja}%
  \BibitemOpen
  \bibfield  {author} {\bibinfo {author} {\bibfnamefont {W.}~\bibnamefont
  {Hampel}} \emph {et~al.} (\bibinfo {collaboration} {GALLEX}),\ }\href
  {\doibase 10.1016/S0370-2693(97)01562-1} {\bibfield  {journal} {\bibinfo
  {journal} {Phys. Lett. B}\ }\textbf {\bibinfo {volume} {420}},\ \bibinfo
  {pages} {114} (\bibinfo {year} {1998})}\BibitemShut {NoStop}%
\bibitem [{\citenamefont {Kaether}\ \emph {et~al.}(2010)\citenamefont
  {Kaether}, \citenamefont {Hampel}, \citenamefont {Heusser}, \citenamefont
  {Kiko},\ and\ \citenamefont {Kirsten}}]{Kaether:2010ag}%
  \BibitemOpen
  \bibfield  {author} {\bibinfo {author} {\bibfnamefont {F.}~\bibnamefont
  {Kaether}}, \bibinfo {author} {\bibfnamefont {W.}~\bibnamefont {Hampel}},
  \bibinfo {author} {\bibfnamefont {G.}~\bibnamefont {Heusser}}, \bibinfo
  {author} {\bibfnamefont {J.}~\bibnamefont {Kiko}}, \ and\ \bibinfo {author}
  {\bibfnamefont {T.}~\bibnamefont {Kirsten}},\ }\href {\doibase
  10.1016/j.physletb.2010.01.030} {\bibfield  {journal} {\bibinfo  {journal}
  {Phys. Lett. B}\ }\textbf {\bibinfo {volume} {685}},\ \bibinfo {pages} {47}
  (\bibinfo {year} {2010})},\ \Eprint {http://arxiv.org/abs/1001.2731}
  {arXiv:1001.2731 [hep-ex]} \BibitemShut {NoStop}%
\bibitem [{\citenamefont {Abdurashitov}\ \emph {et~al.}(1996)\citenamefont
  {Abdurashitov} \emph {et~al.}}]{Abdurashitov:1996dp}%
  \BibitemOpen
  \bibfield  {author} {\bibinfo {author} {\bibfnamefont {D.~N.}\ \bibnamefont
  {Abdurashitov}} \emph {et~al.},\ }\href {\doibase
  10.1103/PhysRevLett.77.4708} {\bibfield  {journal} {\bibinfo  {journal}
  {Phys. Rev. Lett.}\ }\textbf {\bibinfo {volume} {77}},\ \bibinfo {pages}
  {4708} (\bibinfo {year} {1996})}\BibitemShut {NoStop}%
\bibitem [{\citenamefont {Giunti}\ and\ \citenamefont
  {Laveder}(2011)}]{Giunti:2010zu}%
  \BibitemOpen
  \bibfield  {author} {\bibinfo {author} {\bibfnamefont {C.}~\bibnamefont
  {Giunti}}\ and\ \bibinfo {author} {\bibfnamefont {M.}~\bibnamefont
  {Laveder}},\ }\href {\doibase 10.1103/PhysRevC.83.065504} {\bibfield
  {journal} {\bibinfo  {journal} {Phys. Rev. C}\ }\textbf {\bibinfo {volume}
  {83}},\ \bibinfo {pages} {065504} (\bibinfo {year} {2011})},\ \Eprint
  {http://arxiv.org/abs/1006.3244} {arXiv:1006.3244 [hep-ph]} \BibitemShut
  {NoStop}%
\bibitem [{\citenamefont {Kostensalo}\ \emph {et~al.}(2019)\citenamefont
  {Kostensalo}, \citenamefont {Suhonen}, \citenamefont {Giunti},\ and\
  \citenamefont {Srivastava}}]{Kostensalo:2019vmv}%
  \BibitemOpen
  \bibfield  {author} {\bibinfo {author} {\bibfnamefont {J.}~\bibnamefont
  {Kostensalo}}, \bibinfo {author} {\bibfnamefont {J.}~\bibnamefont {Suhonen}},
  \bibinfo {author} {\bibfnamefont {C.}~\bibnamefont {Giunti}}, \ and\ \bibinfo
  {author} {\bibfnamefont {P.~C.}\ \bibnamefont {Srivastava}},\ }\href
  {\doibase 10.1016/j.physletb.2019.06.057} {\bibfield  {journal} {\bibinfo
  {journal} {Phys. Lett. B}\ }\textbf {\bibinfo {volume} {795}},\ \bibinfo
  {pages} {542} (\bibinfo {year} {2019})},\ \Eprint
  {http://arxiv.org/abs/1906.10980} {arXiv:1906.10980 [nucl-th]} \BibitemShut
  {NoStop}%
\bibitem [{\citenamefont {Barinov}\ \emph {et~al.}(2022)\citenamefont {Barinov}
  \emph {et~al.}}]{Barinov:2021asz}%
  \BibitemOpen
  \bibfield  {author} {\bibinfo {author} {\bibfnamefont {V.~V.}\ \bibnamefont
  {Barinov}} \emph {et~al.},\ }\href {\doibase 10.1103/PhysRevLett.128.232501}
  {\bibfield  {journal} {\bibinfo  {journal} {Phys. Rev. Lett.}\ }\textbf
  {\bibinfo {volume} {128}},\ \bibinfo {pages} {232501} (\bibinfo {year}
  {2022})},\ \Eprint {http://arxiv.org/abs/2109.11482} {arXiv:2109.11482
  [nucl-ex]} \BibitemShut {NoStop}%
\bibitem [{\citenamefont {Parke}\ and\ \citenamefont
  {Ross-Lonergan}(2016)}]{Parke:2015goa}%
  \BibitemOpen
  \bibfield  {author} {\bibinfo {author} {\bibfnamefont {S.}~\bibnamefont
  {Parke}}\ and\ \bibinfo {author} {\bibfnamefont {M.}~\bibnamefont
  {Ross-Lonergan}},\ }\href {\doibase 10.1103/PhysRevD.93.113009} {\bibfield
  {journal} {\bibinfo  {journal} {Phys. Rev. D}\ }\textbf {\bibinfo {volume}
  {93}},\ \bibinfo {pages} {113009} (\bibinfo {year} {2016})},\ \Eprint
  {http://arxiv.org/abs/1508.05095} {arXiv:1508.05095 [hep-ph]} \BibitemShut
  {NoStop}%
\bibitem [{\citenamefont {Ellis}\ \emph {et~al.}(2020)\citenamefont {Ellis},
  \citenamefont {Kelly},\ and\ \citenamefont {Li}}]{Ellis:2020hus}%
  \BibitemOpen
  \bibfield  {author} {\bibinfo {author} {\bibfnamefont {S.~A.~R.}\
  \bibnamefont {Ellis}}, \bibinfo {author} {\bibfnamefont {K.~J.}\ \bibnamefont
  {Kelly}}, \ and\ \bibinfo {author} {\bibfnamefont {S.~W.}\ \bibnamefont
  {Li}},\ }\href {\doibase 10.1007/JHEP12(2020)068} {\bibfield  {journal}
  {\bibinfo  {journal} {JHEP}\ }\textbf {\bibinfo {volume} {12}},\ \bibinfo
  {pages} {068} (\bibinfo {year} {2020})},\ \Eprint
  {http://arxiv.org/abs/2008.01088} {arXiv:2008.01088 [hep-ph]} \BibitemShut
  {NoStop}%
\bibitem [{\citenamefont {Hu}\ \emph {et~al.}(2021)\citenamefont {Hu},
  \citenamefont {Ling}, \citenamefont {Tang},\ and\ \citenamefont
  {Wang}}]{Hu:2020oba}%
  \BibitemOpen
  \bibfield  {author} {\bibinfo {author} {\bibfnamefont {Z.}~\bibnamefont
  {Hu}}, \bibinfo {author} {\bibfnamefont {J.}~\bibnamefont {Ling}}, \bibinfo
  {author} {\bibfnamefont {J.}~\bibnamefont {Tang}}, \ and\ \bibinfo {author}
  {\bibfnamefont {T.}~\bibnamefont {Wang}},\ }\href {\doibase
  10.1007/JHEP01(2021)124} {\bibfield  {journal} {\bibinfo  {journal} {JHEP}\
  }\textbf {\bibinfo {volume} {01}},\ \bibinfo {pages} {124} (\bibinfo {year}
  {2021})},\ \Eprint {http://arxiv.org/abs/2008.09730} {arXiv:2008.09730
  [hep-ph]} \BibitemShut {NoStop}%
\bibitem [{\citenamefont {Bolton}\ \emph {et~al.}(2020)\citenamefont {Bolton},
  \citenamefont {Deppisch},\ and\ \citenamefont {Bhupal~Dev}}]{Bolton:2019pcu}%
  \BibitemOpen
  \bibfield  {author} {\bibinfo {author} {\bibfnamefont {P.~D.}\ \bibnamefont
  {Bolton}}, \bibinfo {author} {\bibfnamefont {F.~F.}\ \bibnamefont
  {Deppisch}}, \ and\ \bibinfo {author} {\bibfnamefont {P.~S.}\ \bibnamefont
  {Bhupal~Dev}},\ }\href {\doibase 10.1007/JHEP03(2020)170} {\bibfield
  {journal} {\bibinfo  {journal} {JHEP}\ }\textbf {\bibinfo {volume} {03}},\
  \bibinfo {pages} {170} (\bibinfo {year} {2020})},\ \Eprint
  {http://arxiv.org/abs/1912.03058} {arXiv:1912.03058 [hep-ph]} \BibitemShut
  {NoStop}%
\bibitem [{\citenamefont {Bergsma}\ \emph {et~al.}(1986)\citenamefont {Bergsma}
  \emph {et~al.}}]{CHARM:1985nku}%
  \BibitemOpen
  \bibfield  {author} {\bibinfo {author} {\bibfnamefont {F.}~\bibnamefont
  {Bergsma}} \emph {et~al.} (\bibinfo {collaboration} {CHARM}),\ }\href
  {\doibase 10.1016/0370-2693(86)91601-1} {\bibfield  {journal} {\bibinfo
  {journal} {Phys. Lett. B}\ }\textbf {\bibinfo {volume} {166}},\ \bibinfo
  {pages} {473} (\bibinfo {year} {1986})}\BibitemShut {NoStop}%
\bibitem [{\citenamefont {Vilain}\ \emph {et~al.}(1995)\citenamefont {Vilain}
  \emph {et~al.}}]{CHARMII:1994jjr}%
  \BibitemOpen
  \bibfield  {author} {\bibinfo {author} {\bibfnamefont {P.}~\bibnamefont
  {Vilain}} \emph {et~al.} (\bibinfo {collaboration} {CHARM II}),\ }\href
  {\doibase 10.1016/0370-2693(94)01422-9} {\bibfield  {journal} {\bibinfo
  {journal} {Phys. Lett. B}\ }\textbf {\bibinfo {volume} {343}},\ \bibinfo
  {pages} {453} (\bibinfo {year} {1995})}\BibitemShut {NoStop}%
\bibitem [{\citenamefont {Abe}\ \emph {et~al.}(2019)\citenamefont {Abe} \emph
  {et~al.}}]{T2K:2019jwa}%
  \BibitemOpen
  \bibfield  {author} {\bibinfo {author} {\bibfnamefont {K.}~\bibnamefont
  {Abe}} \emph {et~al.} (\bibinfo {collaboration} {T2K}),\ }\href {\doibase
  10.1103/PhysRevD.100.052006} {\bibfield  {journal} {\bibinfo  {journal}
  {Phys. Rev. D}\ }\textbf {\bibinfo {volume} {100}},\ \bibinfo {pages}
  {052006} (\bibinfo {year} {2019})},\ \Eprint
  {http://arxiv.org/abs/1902.07598} {arXiv:1902.07598 [hep-ex]} \BibitemShut
  {NoStop}%
\bibitem [{\citenamefont {Bryman}\ and\ \citenamefont
  {Shrock}(2019)}]{Bryman:2019bjg}%
  \BibitemOpen
  \bibfield  {author} {\bibinfo {author} {\bibfnamefont {D.~A.}\ \bibnamefont
  {Bryman}}\ and\ \bibinfo {author} {\bibfnamefont {R.}~\bibnamefont
  {Shrock}},\ }\href {\doibase 10.1103/PhysRevD.100.073011} {\bibfield
  {journal} {\bibinfo  {journal} {Phys. Rev. D}\ }\textbf {\bibinfo {volume}
  {100}},\ \bibinfo {pages} {073011} (\bibinfo {year} {2019})},\ \Eprint
  {http://arxiv.org/abs/1909.11198} {arXiv:1909.11198 [hep-ph]} \BibitemShut
  {NoStop}%
\bibitem [{\citenamefont {Bellini}\ \emph {et~al.}(2013)\citenamefont {Bellini}
  \emph {et~al.}}]{Borexino:2013bot}%
  \BibitemOpen
  \bibfield  {author} {\bibinfo {author} {\bibfnamefont {G.}~\bibnamefont
  {Bellini}} \emph {et~al.} (\bibinfo {collaboration} {Borexino}),\ }\href
  {\doibase 10.1103/PhysRevD.88.072010} {\bibfield  {journal} {\bibinfo
  {journal} {Phys. Rev. D}\ }\textbf {\bibinfo {volume} {88}},\ \bibinfo
  {pages} {072010} (\bibinfo {year} {2013})},\ \Eprint
  {http://arxiv.org/abs/1311.5347} {arXiv:1311.5347 [hep-ex]} \BibitemShut
  {NoStop}%
\bibitem [{\citenamefont {Ruchayskiy}\ and\ \citenamefont
  {Ivashko}(2012)}]{Ruchayskiy:2012si}%
  \BibitemOpen
  \bibfield  {author} {\bibinfo {author} {\bibfnamefont {O.}~\bibnamefont
  {Ruchayskiy}}\ and\ \bibinfo {author} {\bibfnamefont {A.}~\bibnamefont
  {Ivashko}},\ }\href {\doibase 10.1088/1475-7516/2012/10/014} {\bibfield
  {journal} {\bibinfo  {journal} {JCAP}\ }\textbf {\bibinfo {volume} {10}},\
  \bibinfo {pages} {014} (\bibinfo {year} {2012})},\ \Eprint
  {http://arxiv.org/abs/1202.2841} {arXiv:1202.2841 [hep-ph]} \BibitemShut
  {NoStop}%
\bibitem [{\citenamefont {Vincent}\ \emph {et~al.}(2015)\citenamefont
  {Vincent}, \citenamefont {Martinez}, \citenamefont {Hern\'andez},
  \citenamefont {Lattanzi},\ and\ \citenamefont {Mena}}]{Vincent:2014rja}%
  \BibitemOpen
  \bibfield  {author} {\bibinfo {author} {\bibfnamefont {A.~C.}\ \bibnamefont
  {Vincent}}, \bibinfo {author} {\bibfnamefont {E.~F.}\ \bibnamefont
  {Martinez}}, \bibinfo {author} {\bibfnamefont {P.}~\bibnamefont
  {Hern\'andez}}, \bibinfo {author} {\bibfnamefont {M.}~\bibnamefont
  {Lattanzi}}, \ and\ \bibinfo {author} {\bibfnamefont {O.}~\bibnamefont
  {Mena}},\ }\href {\doibase 10.1088/1475-7516/2015/04/006} {\bibfield
  {journal} {\bibinfo  {journal} {JCAP}\ }\textbf {\bibinfo {volume} {04}},\
  \bibinfo {pages} {006} (\bibinfo {year} {2015})},\ \Eprint
  {http://arxiv.org/abs/1408.1956} {arXiv:1408.1956 [astro-ph.CO]} \BibitemShut
  {NoStop}%
\bibitem [{\citenamefont {Davoudiasl}\ and\ \citenamefont
  {Denton}(2023)}]{Davoudiasl:2023uiq}%
  \BibitemOpen
  \bibfield  {author} {\bibinfo {author} {\bibfnamefont {H.}~\bibnamefont
  {Davoudiasl}}\ and\ \bibinfo {author} {\bibfnamefont {P.~B.}\ \bibnamefont
  {Denton}},\ }\href@noop {} {\  (\bibinfo {year} {2023})},\ \Eprint
  {http://arxiv.org/abs/2301.09651} {arXiv:2301.09651 [hep-ph]} \BibitemShut
  {NoStop}%
\bibitem [{\citenamefont {Brdar}\ \emph {et~al.}(2023)\citenamefont {Brdar},
  \citenamefont {Gehrlein},\ and\ \citenamefont {Kopp}}]{Brdar:2023cms}%
  \BibitemOpen
  \bibfield  {author} {\bibinfo {author} {\bibfnamefont {V.}~\bibnamefont
  {Brdar}}, \bibinfo {author} {\bibfnamefont {J.}~\bibnamefont {Gehrlein}}, \
  and\ \bibinfo {author} {\bibfnamefont {J.}~\bibnamefont {Kopp}},\ }\href
  {\doibase 10.1007/JHEP05(2023)143} {\bibfield  {journal} {\bibinfo  {journal}
  {JHEP}\ }\textbf {\bibinfo {volume} {05}},\ \bibinfo {pages} {143} (\bibinfo
  {year} {2023})},\ \Eprint {http://arxiv.org/abs/2303.05528} {arXiv:2303.05528
  [hep-ph]} \BibitemShut {NoStop}%
\bibitem [{\citenamefont {Gu}\ and\ \citenamefont {He}(2017)}]{Gu:2016ege}%
  \BibitemOpen
  \bibfield  {author} {\bibinfo {author} {\bibfnamefont {P.-H.}\ \bibnamefont
  {Gu}}\ and\ \bibinfo {author} {\bibfnamefont {X.-G.}\ \bibnamefont {He}},\
  }\href {\doibase 10.1016/j.nuclphysb.2017.03.023} {\bibfield  {journal}
  {\bibinfo  {journal} {Nucl. Phys. B}\ }\textbf {\bibinfo {volume} {919}},\
  \bibinfo {pages} {209} (\bibinfo {year} {2017})},\ \Eprint
  {http://arxiv.org/abs/1606.05171} {arXiv:1606.05171 [hep-ph]} \BibitemShut
  {NoStop}%
\bibitem [{\citenamefont {Barbieri}\ and\ \citenamefont
  {Ericson}(1975)}]{Barbieri:1975xy}%
  \BibitemOpen
  \bibfield  {author} {\bibinfo {author} {\bibfnamefont {R.}~\bibnamefont
  {Barbieri}}\ and\ \bibinfo {author} {\bibfnamefont {T.~E.~O.}\ \bibnamefont
  {Ericson}},\ }\href {\doibase 10.1016/0370-2693(75)90073-8} {\bibfield
  {journal} {\bibinfo  {journal} {Phys. Lett. B}\ }\textbf {\bibinfo {volume}
  {57}},\ \bibinfo {pages} {270} (\bibinfo {year} {1975})}\BibitemShut
  {NoStop}%
\bibitem [{\citenamefont {Bordes}\ \emph {et~al.}(2019)\citenamefont {Bordes},
  \citenamefont {Chan},\ and\ \citenamefont {Tsou}}]{Bordes:2019qjk}%
  \BibitemOpen
  \bibfield  {author} {\bibinfo {author} {\bibfnamefont {J.}~\bibnamefont
  {Bordes}}, \bibinfo {author} {\bibfnamefont {H.-M.}\ \bibnamefont {Chan}}, \
  and\ \bibinfo {author} {\bibfnamefont {S.~T.}\ \bibnamefont {Tsou}},\ }\href
  {\doibase 10.1142/S0217751X19501409} {\bibfield  {journal} {\bibinfo
  {journal} {Int. J. Mod. Phys. A}\ }\textbf {\bibinfo {volume} {34}},\
  \bibinfo {pages} {1950140} (\bibinfo {year} {2019})},\ \Eprint
  {http://arxiv.org/abs/1906.09229} {arXiv:1906.09229 [hep-ph]} \BibitemShut
  {NoStop}%
\bibitem [{\citenamefont {Nam}(2020)}]{Nam:2019osu}%
  \BibitemOpen
  \bibfield  {author} {\bibinfo {author} {\bibfnamefont {C.~H.}\ \bibnamefont
  {Nam}},\ }\href {\doibase 10.1140/epjc/s10052-020-7794-0} {\bibfield
  {journal} {\bibinfo  {journal} {Eur. Phys. J. C}\ }\textbf {\bibinfo {volume}
  {80}},\ \bibinfo {pages} {231} (\bibinfo {year} {2020})},\ \Eprint
  {http://arxiv.org/abs/1907.09819} {arXiv:1907.09819 [hep-ph]} \BibitemShut
  {NoStop}%
\bibitem [{\citenamefont {Depero}\ \emph {et~al.}(2020)\citenamefont {Depero}
  \emph {et~al.}}]{NA64:2020xxh}%
  \BibitemOpen
  \bibfield  {author} {\bibinfo {author} {\bibfnamefont {E.}~\bibnamefont
  {Depero}} \emph {et~al.} (\bibinfo {collaboration} {NA64}),\ }\href {\doibase
  10.1140/epjc/s10052-020-08725-x} {\bibfield  {journal} {\bibinfo  {journal}
  {Eur. Phys. J. C}\ }\textbf {\bibinfo {volume} {80}},\ \bibinfo {pages}
  {1159} (\bibinfo {year} {2020})},\ \Eprint {http://arxiv.org/abs/2009.02756}
  {arXiv:2009.02756 [hep-ex]} \BibitemShut {NoStop}%
\bibitem [{\citenamefont {Chen}\ \emph {et~al.}(2016)\citenamefont {Chen},
  \citenamefont {Liang},\ and\ \citenamefont {Qiao}}]{Chen:2016dhm}%
  \BibitemOpen
  \bibfield  {author} {\bibinfo {author} {\bibfnamefont {L.-B.}\ \bibnamefont
  {Chen}}, \bibinfo {author} {\bibfnamefont {Y.}~\bibnamefont {Liang}}, \ and\
  \bibinfo {author} {\bibfnamefont {C.-F.}\ \bibnamefont {Qiao}},\ }\href@noop
  {} {\  (\bibinfo {year} {2016})},\ \Eprint {http://arxiv.org/abs/1607.03970}
  {arXiv:1607.03970 [hep-ph]} \BibitemShut {NoStop}%
\bibitem [{\citenamefont {Chiang}\ and\ \citenamefont
  {Tseng}(2017)}]{Chiang:2016cyf}%
  \BibitemOpen
  \bibfield  {author} {\bibinfo {author} {\bibfnamefont {C.-W.}\ \bibnamefont
  {Chiang}}\ and\ \bibinfo {author} {\bibfnamefont {P.-Y.}\ \bibnamefont
  {Tseng}},\ }\href {\doibase 10.1016/j.physletb.2017.02.022} {\bibfield
  {journal} {\bibinfo  {journal} {Phys. Lett. B}\ }\textbf {\bibinfo {volume}
  {767}},\ \bibinfo {pages} {289} (\bibinfo {year} {2017})},\ \Eprint
  {http://arxiv.org/abs/1612.06985} {arXiv:1612.06985 [hep-ph]} \BibitemShut
  {NoStop}%
\bibitem [{\citenamefont {Azuelos}\ \emph {et~al.}(2022)\citenamefont {Azuelos}
  \emph {et~al.}}]{Azuelos:2022nbu}%
  \BibitemOpen
  \bibfield  {author} {\bibinfo {author} {\bibfnamefont {G.}~\bibnamefont
  {Azuelos}} \emph {et~al.},\ }\href {\doibase 10.1088/1742-6596/2391/1/012008}
  {\bibfield  {journal} {\bibinfo  {journal} {J. Phys. Conf. Ser.}\ }\textbf
  {\bibinfo {volume} {2391}},\ \bibinfo {pages} {012008} (\bibinfo {year}
  {2022})},\ \Eprint {http://arxiv.org/abs/2211.11900} {arXiv:2211.11900
  [physics.ins-det]} \BibitemShut {NoStop}%
\bibitem [{\citenamefont {Cline}\ \emph {et~al.}(2022)\citenamefont {Cline}
  \emph {et~al.}}]{DarkLight:2022uji}%
  \BibitemOpen
  \bibfield  {author} {\bibinfo {author} {\bibfnamefont {E.}~\bibnamefont
  {Cline}} \emph {et~al.} (\bibinfo {collaboration} {DarkLight}),\ }\href
  {\doibase 10.1088/1742-6596/2391/1/012010} {\bibfield  {journal} {\bibinfo
  {journal} {J. Phys. Conf. Ser.}\ }\textbf {\bibinfo {volume} {2391}},\
  \bibinfo {pages} {012010} (\bibinfo {year} {2022})},\ \Eprint
  {http://arxiv.org/abs/2208.04120} {arXiv:2208.04120 [nucl-ex]} \BibitemShut
  {NoStop}%
\bibitem [{\citenamefont {Dutta}\ \emph {et~al.}(2023)\citenamefont {Dutta}
  \emph {et~al.}}]{Dutta:2023ifr}%
  \BibitemOpen
  \bibfield  {author} {\bibinfo {author} {\bibfnamefont {D.}~\bibnamefont
  {Dutta}} \emph {et~al.},\ }\href@noop {} {\  (\bibinfo {year} {2023})},\
  \Eprint {http://arxiv.org/abs/2301.08768} {arXiv:2301.08768 [nucl-ex]}
  \BibitemShut {NoStop}%
\bibitem [{\citenamefont {Darm\'e}\ \emph {et~al.}(2022)\citenamefont
  {Darm\'e}, \citenamefont {Mancini}, \citenamefont {Nardi},\ and\
  \citenamefont {Raggi}}]{Darme:2022zfw}%
  \BibitemOpen
  \bibfield  {author} {\bibinfo {author} {\bibfnamefont {L.}~\bibnamefont
  {Darm\'e}}, \bibinfo {author} {\bibfnamefont {M.}~\bibnamefont {Mancini}},
  \bibinfo {author} {\bibfnamefont {E.}~\bibnamefont {Nardi}}, \ and\ \bibinfo
  {author} {\bibfnamefont {M.}~\bibnamefont {Raggi}},\ }\href {\doibase
  10.1103/PhysRevD.106.115036} {\bibfield  {journal} {\bibinfo  {journal}
  {Phys. Rev. D}\ }\textbf {\bibinfo {volume} {106}},\ \bibinfo {pages}
  {115036} (\bibinfo {year} {2022})},\ \Eprint
  {http://arxiv.org/abs/2209.09261} {arXiv:2209.09261 [hep-ph]} \BibitemShut
  {NoStop}%
\bibitem [{\citenamefont {Caricato}\ \emph {et~al.}(2022)\citenamefont
  {Caricato} \emph {et~al.}}]{PADME:2022xly}%
  \BibitemOpen
  \bibfield  {author} {\bibinfo {author} {\bibfnamefont {A.~P.}\ \bibnamefont
  {Caricato}} \emph {et~al.} (\bibinfo {collaboration} {PADME}),\ }\href@noop
  {} {\  (\bibinfo {year} {2022})},\ \Eprint {http://arxiv.org/abs/2209.14755}
  {arXiv:2209.14755 [hep-ex]} \BibitemShut {NoStop}%
\bibitem [{\citenamefont {Andreev}\ \emph {et~al.}(2021)\citenamefont {Andreev}
  \emph {et~al.}}]{NA64:2021xzo}%
  \BibitemOpen
  \bibfield  {author} {\bibinfo {author} {\bibfnamefont {Y.~M.}\ \bibnamefont
  {Andreev}} \emph {et~al.} (\bibinfo {collaboration} {NA64}),\ }\href
  {\doibase 10.1103/PhysRevLett.126.211802} {\bibfield  {journal} {\bibinfo
  {journal} {Phys. Rev. Lett.}\ }\textbf {\bibinfo {volume} {126}},\ \bibinfo
  {pages} {211802} (\bibinfo {year} {2021})},\ \Eprint
  {http://arxiv.org/abs/2102.01885} {arXiv:2102.01885 [hep-ex]} \BibitemShut
  {NoStop}%
\bibitem [{\citenamefont {Deniz}\ \emph {et~al.}(2010)\citenamefont {Deniz}
  \emph {et~al.}}]{TEXONO:2009knm}%
  \BibitemOpen
  \bibfield  {author} {\bibinfo {author} {\bibfnamefont {M.}~\bibnamefont
  {Deniz}} \emph {et~al.} (\bibinfo {collaboration} {TEXONO}),\ }\href
  {\doibase 10.1103/PhysRevD.81.072001} {\bibfield  {journal} {\bibinfo
  {journal} {Phys. Rev. D}\ }\textbf {\bibinfo {volume} {81}},\ \bibinfo
  {pages} {072001} (\bibinfo {year} {2010})},\ \Eprint
  {http://arxiv.org/abs/0911.1597} {arXiv:0911.1597 [hep-ex]} \BibitemShut
  {NoStop}%
\bibitem [{\citenamefont {Suliga}\ \emph {et~al.}(2021)\citenamefont {Suliga},
  \citenamefont {Shalgar},\ and\ \citenamefont {Fuller}}]{Suliga:2020lir}%
  \BibitemOpen
  \bibfield  {author} {\bibinfo {author} {\bibfnamefont {A.~M.}\ \bibnamefont
  {Suliga}}, \bibinfo {author} {\bibfnamefont {S.}~\bibnamefont {Shalgar}}, \
  and\ \bibinfo {author} {\bibfnamefont {G.~M.}\ \bibnamefont {Fuller}},\
  }\href {\doibase 10.1088/1475-7516/2021/07/042} {\bibfield  {journal}
  {\bibinfo  {journal} {JCAP}\ }\textbf {\bibinfo {volume} {07}},\ \bibinfo
  {pages} {042} (\bibinfo {year} {2021})},\ \Eprint
  {http://arxiv.org/abs/2012.11620} {arXiv:2012.11620 [astro-ph.HE]}
  \BibitemShut {NoStop}%
\end{thebibliography}%

\end{document}